\title{Informative Imitation\footnote{We thank Aislinn Bohren, Marina Halac, Kevin He, Moritz Meyer-ter-Vehn, Bruno Strulovici, and Alexander Wolitzky for helpful comments. Pei thanks the NSF Grant SES-2337566 for financial support.}}
\author{Nemanja Anti\'{c}\footnote{Northwestern University, \url{nemanja.antic@kellogg.northwestern.edu}} \and Harry Pei\footnote{Northwestern University, \url{harrydp@northwestern.edu}}}
\date{\today}
\documentclass[pstricks,12pt]{article}
\usepackage{latexsym}
\usepackage{graphicx}
\usepackage{epsfig}
\usepackage{dsfont}
\usepackage{color}
\usepackage{natbib}
\usepackage{epstopdf}
\usepackage{hyperref}
\hypersetup{
    colorlinks=true,
    linkcolor=blue,
    citecolor=blue,
    filecolor=blue,
    urlcolor=blue
}
\usepackage{comment}
\usepackage[T1]{fontenc}
\usepackage{amsmath, amsthm, amssymb}
\usepackage{pstricks, pst-node}
\usepackage{tikz}
\usepackage{caption}
\usepackage{graphicx}
\graphicspath{ {./figures/} }
\usepackage{tikz}
\usetikzlibrary{decorations.pathreplacing, calc}
\usepackage{pgfplots}
\usepackage{subcaption}
\pgfplotsset{compat=1.18}
\pgfplotsset{
  every axis/.append style={font=\small}
}
\tikzset{declare function={
  LA(\xx,\mm,\pp,\gg)=((1-(1-\pp)*\gg)/(1-\gg))^\mm*(1-(1-\pp)*\xx*(1-(1-\gg)^\mm))/(1-\xx*(1-(1-(1-\pp)*\gg)^\mm));
  LU(\xx,\mm,\pp,\gg)=((1-(1-\pp)*\gg)/(1-\gg))^\mm*(\pp*(1-\pp)*\xx*(1-\gg)^\mm+(1-(1-\pp)*\xx)*(\pp+(1-\pp)*\gg^\mm))/(\pp*\xx*(1-(1-\pp)*\gg)^\mm+(1-\xx)*(\pp+(1-\pp)^(\mm+1)*(1-\gg)*\gg^\mm));
  ELL(\Lzero,\Lone,\R)=ifthenelse(\R<=\Lone,\Lone,min(\R,\Lzero));
}}
\usepackage{enumitem}

\newlength{\lemmaheadwidth}
\makeatletter
\newcommand{\measurelemmahead}{%
  \setlength{\lemmaheadwidth}{\wd\@labels}%
}
\makeatother

\definecolor{figureAware}{RGB}{0,114,178}
\definecolor{figureUnaware}{RGB}{213,94,0}

\begin{document}
\numberwithin{equation}{section}

\maketitle
\thispagestyle{empty}

\begin{abstract}
We study steady-state equilibria of a sequential learning game in which each agent chooses between a safe action and a risky action of unknown profitability. Before choosing, each agent observes how many of the last $m$ agents adopted the risky action, but not whether they were informed or what their adoption costs were. We show that uninformed agents' imitation decisions can be strategic complements: the more likely uninformed agents are to join risky-action herds, the more informative those herds become about the action's profitability. When this complementarity gives rise to multiple equilibria, the one that maximizes the probability of imitation by uninformed agents also maximizes agents' welfare.

\noindent \textbf{Keywords:} social learning, steady state, informativeness, herding.  

    \noindent \textbf{JEL Codes:} C72, C73, D82, D83.
\end{abstract}

\newtheorem{Proposition}{\hskip\parindent\bf{Proposition}}
\newtheorem{Theorem}{\hskip\parindent\bf{Theorem}}
\newtheorem{Lemma}{\hskip\parindent\bf{Lemma}}
\newtheorem{Corollary}{\hskip\parindent\bf{Corollary}}
\newtheorem*{Definition}{\hskip\parindent\bf{Threshold Strategy}}
\newtheorem{Assumption}{\hskip\parindent\bf{Assumption}}
\newtheorem{Condition}{\hskip\parindent\bf{Condition}}
\newtheorem{Claim}{\hskip\parindent\bf{Claim}}

\addtocounter{page}{-1}

\clearpage

\section{Introduction}\label{sec1}
People sometimes imitate others’ choices without fully understanding the rationales behind them. Diners tend to order popular dishes in restaurants without knowing how informed previous diners were \citep*{cai2009observational}. Investors are more likely to purchase financial assets that their peers have bought, even without knowing how informed those peers were or what opportunity costs they faced \citep*{bursztyn2014understanding}. Similarly, music listeners are influenced by the songs others have downloaded, even when they cannot observe the factors underlying those choices \citep*{salganik2006experimental}.

This paper studies a sequential learning model in which agents with heterogeneous adoption costs and heterogeneous information decide whether to adopt a risky action of unknown profitability. Before deciding, each agent observes the choices of his most recent predecessors, but not their adoption costs or whether they were informed. We show that, even in the absence of payoff externalities, herds can become more informative as uninformed agents become more likely to join them, generating strategic complementarities in uninformed agents' herding decisions.

We analyze steady-state equilibria of a doubly-infinite time horizon game with one active agent in each period. Each agent chooses between a safe action with a known payoff and a risky action, whose profitability, the \textit{state}, is fixed over time and is either \textit{high} or \textit{low}. 
The agent knows the state with some probability, observes his own cost of adopting the risky action, which is either \textit{high} or \textit{low}, and how many of his last $m \in \mathbb{N}$ predecessors took the risky action.\footnote{In this baseline setting where agents have two possible costs, observing the exact sequence of the last $m$ actions yields the same set of equilibrium outcomes as observing only summary statistics (see Proposition \ref{Prop:order}).}
Low-cost agents always prefer the risky action. High-cost agents prefer the risky action only if their posterior of the high state exceeds some threshold, which we assume is above their prior. 

A first observation is that in every equilibrium, uninformed high-cost agents will take the risky action only if all of their last $m$ predecessors did so, i.e., they observe a \textit{herd}. To see why, suppose they take the risky action without observing a herd. Then they would 
do so for sure after observing a herd, making herds absorbing in the high state. Consequently, not observing a herd would reveal the low state, contradicting agents' incentives to take the risky action there.

Our main result is that as long as $m \geq 2$,\footnote{If $m=1$, both in this baseline case and when agents are initially unaware of the risky action, the steady-state informativeness of $1$-agent herds is constant in the probability with which uninformed agents join them.} the steady-state informativeness of herds increases with the probability that uninformed agents join them. Hence, uninformed agents' actions are strategic complements, resulting in a multiplicity of equilibria for an open set of parameters.

The driving force behind our result is that agents cannot tell whether an observed herd is part of a longer sequence of risky actions. Uninformed agents' imitation lengthens such sequences in expectation, making it more likely that an observed herd is part of a longer sequence. This improves the informativeness of herds, because longer sequences of the risky action contain more choices made by informed agents on average and are therefore more indicative of the high state.

We then analyze several variations of our model. First, suppose agents are initially unaware of the risky action: they can take it either when at least one of their last $m$ predecessors did so or when they observe the state themselves.\footnote{Intuitively, since the state reveals the profitability of the risky action, it also reveals the action's existence.}  We show that when $m \geq 2$, whether uninformed agents' herding behaviors are complements or substitutes depends on the type distribution. 

Intuitively, since uninformed low-cost agents cannot take the risky action when they are unaware of it, a risky action taken when none of the last $m$ agents took it is more informative about the state than a risky action taken elsewhere. Imitation may then lower the informativeness of herds, since it raises the chance that uninformed agents are aware of the risky action, which counteracts its effect of increasing the duration of herds. This ``awareness'' effect dominates when (i) a majority of agents have low costs, 
so that a restart of the risky action is
more likely to develop into a herd than to terminate midway, and (ii) most agents are uninformed, in which case uninformed agents can remain unaware of the risky action for a long time.

Next, suppose agents have three possible costs: the \textit{low-cost} agents who always prefer the risky action, the \textit{high-cost} agents whose optimal action depends on the state, and the \textit{highest-cost} agents who always prefer the safe action. 
In this setting, uninformed high-cost agents sometimes take the risky action when they observe herds with size less than $m$, since the lack of risky actions can be partially attributed to some of their predecessors having the highest cost.  
Nevertheless, herding behaviors remain strategic complements under some parameter configurations. We also show that whenever 
such complementarities lead to multiple equilibria, the one that maximizes uninformed high-cost agents' imitation probability also maximizes agents' welfare.

Our paper contributes to the social learning literature, in which most papers either assume that players observe the entire sequence of past actions \citep[e.g.,][]{banerjee1992simple, bikhchandani1992theory, smith2000pathological, kartik2024beyond}, or that they observe a subset of past actions while being able to attribute each observed action to the exact person who took it \citep[e.g.,][]{AcemogluDahlehLobelOzdaglar2011,lobel2015information}. 
\citet{callander2009wisdom}, \citet*{guarino2011aggregate}, and
\citet*{acemoglu2022learning} study models where agents observe the summary statistics of \textit{all} past actions. 
In contrast, the agents in our model can only observe the summary statistics of the last $m$ actions,\footnote{\citet{pei2024reputation} studies a reputation model where each uninformed short-run player observes the number of times an informed long-run player took each of his actions in the last $m$ periods.}  
under which agents who observe $m$-agent herds do not know how long these herds have lasted.

\citet{monzon2014observational},
\citet{banerjee2004word}, and \citet{dasaratha2022learning} study models where players observe only the number of times each action was taken among a randomly sampled set of predecessors, under various sampling rules and sample sizes.\footnote{In \citet{ellison1993rules, ellison1995word}, each agent observes the summary statistics of the actions taken by a random sample of other agents, and then instead of playing a best reply, chooses an action using some heuristics.
\citet{wolitzky2018learning} studies a social learning model similar to \citet{banerjee2004word}'s in which each agent samples $k$ of his predecessors, and observes the \textit{outcomes} of sampled players but not their actions.} We examine a different sampling rule: each agent observes the summary statistics of the actions taken by their last $m$ predecessors. Whereas these papers examine long-run learning outcomes, our focus is on whether uninformed agents' herding behaviors are complements or substitutes.

The closest related paper is \citet{banerjee1993economics}, who studies several models in which agents learn from \textit{rumors}: Each agent observes one immediate predecessor's action but not the predecessor's cost or whether the predecessor knew the state. In his models, agents know exactly when rumors started, under which rational agents follow the rumors only when they receive them early enough. We instead consider settings in which agents do not know when the learning process started. Our steady-state framework fits situations where the risky action (e.g., a technology, a rumor) has existed for a while, making it hard for people to track exactly when it started.

Our paper also relates to a burgeoning literature that analyzes steady-state equilibria of dynamic games such as \citet*{clark2021record} and \citet{antic2026selective}. \citet{acemoglu2014cycles} conduct a steady-state analysis of an overlapping-generations model with payoff externalities and show that 
players' actions are always substitutes. In contrast, we study a model without payoff externalities in which agents' actions are sometimes complements.

Several recent contributions analyze steady-state equilibria of social learning models. For example, \citet{kabos2021welfare} assume that past agents' actions are recorded with noise and examine whether agents put too much weight on their private signals or on others' actions, which differs from our research question. \citet*{dasaratha2023learning} and \citet*{levy2024stationary} assume that agents have homogeneous preferences but that the state evolves over time, which contrasts with our model where agents have heterogeneous preferences.\footnote{Our finding that uninformed agents' imitation improves the informativeness of herds continues to hold when the state is changing over time but is sufficiently persistent.}

\section{Baseline Model}\label{sec2}
Consider a game with a doubly-infinite time horizon $t \in \mathbb{Z} \equiv \{\ldots, -1, 0,1, \ldots \}$, one active agent in each period, and a constant 
state of the world $\theta \in \Theta = \{\underline{\theta},\overline{\theta}\} \subset \mathbb{R}_+$. Let $\pi_0 \in (0,1)$ denote the prior probability of the \textit{high state} $\overline{\theta}$. 
In period $t$, agent $t$ chooses $a_t \in \{0,1\}$, where $a_t = 0$ corresponds to taking a \textit{safe action} and $a_t =1$ to adopting a \textit{risky action}.

Before choosing $a_t$, agent $t$
observes (i) his own cost of adoption, $c_t \in \{c^*, \underline{c}\} \subset \mathbb{R}_+$, with $\underline{c} < c^*$; (ii) the number $n_t := \sum_{k=t-m}^{t-1} a_k \in \{0, 1, \ldots, m\}$ of his $m$ immediate predecessors who adopted, where $m \in \mathbb{Z}_{++}$ is an exogenous parameter,\footnote{In the baseline model, agents cannot observe the exact sequence of their last $m$ predecessors' actions. In Section \ref{sec5}, we will comment on the robustness of our result when agents can observe the exact sequence.} and (iii) with probability $p \in (0,1)$, the realized state $\theta$. 
The costs $\{c_t\}_{t \in \mathbb{Z}}$ are i.i.d. and are independent of $\theta$. Let $\gamma \in (0,1)$ denote the probability that $c_t=c^*$, and let $\underline{\gamma} = 1 - \gamma$ denote the probability
that $c_t = \underline{c}$. Following \citet{banerjee1993economics}, we consider the following two \textit{awareness cases}, indexed by $j \in \{a,u\}$: 
\begin{enumerate}
    \item Agents are \textit{aware} of the risky action (i.e., $j=a$) and can always choose either action. 
    \item Agents are initially \textit{unaware} of the risky action (i.e., $j=u$). They can adopt only if they observe the state $\theta$ or if
    at least one of their $m$ immediate predecessors adopted.
\end{enumerate}

Agent $t$'s payoff is $ (\theta-c_t) \cdot a_t$. Hence, his payoff from taking the safe action is normalized to $0$ and his payoff from adopting the risky action equals $\theta$ minus his cost $c_t$. We assume that
\begin{equation}\label{ass:cost}
\underline{c} <\underline{\theta} \quad  \textrm{and} \quad 
\pi_0 \overline{\theta} + (1-\pi_0) \underline{\theta} < c^* < \overline{\theta}.
\end{equation}
Thus, cost-$\underline{c}$ agents strictly prefer to adopt in both states, and cost-$c^*$ agents prefer to adopt only when their belief that $\theta = \overline{\theta}$ is more than $\pi_* := \frac{c^*-\underline{\theta}}{\overline{\theta}-\underline{\theta}}$, where $\pi_* \in (\pi_0,1)$ by (\ref{ass:cost}).

Our solution concept is steady-state equilibrium (or \textit{equilibrium}). 
In any steady state, the distribution of $n_t$ is independent of $t$, so we let $n$ denote, for a generic agent, the number of people who adopted among his   $m$ immediate predecessors.
An equilibrium consists of (i) a strategy $\widetilde{\sigma}$ that maps an agent's realized cost, whether and, if so, which state he observes, and the value of $n$, into his adoption probability;\footnote{When agents are automatically aware of action $1$, all strategies are feasible. When agents are initially unaware of action $1$, a strategy is feasible if and only if uninformed agents of all costs can only take action $0$ when $n=0$.}
(ii) two distributions $\overline{\mu},\underline{\mu}\in\Delta(\{0,1\}^m)$ over the ordered sequences of $m$ actions conditional on $\overline{\theta}$ and $\underline{\theta}$, respectively; and (iii) a belief map $\pi:\{0,\ldots,m\}\to[0,1]$ that specifies an agent's belief after observing $n$ but not $\theta$. These components satisfy (i) $\widetilde{\sigma}$ is optimal given $\pi$; (ii) $\pi$ is derived from $\overline{\mu}$ and $\underline{\mu}$ according to Bayes' rule; and (iii) $\overline{\mu}$ and $\underline{\mu}$ are the steady-state distributions induced by $\widetilde{\sigma}$ conditional on $\overline{\theta}$ and $\underline{\theta}$, respectively.

Since agents who observe $\theta$ or have cost $\underline{c}$ have strictly dominant actions, it is without loss to focus on the component of $\widetilde{\sigma}$ for uninformed high-cost agents, denoted by $\sigma:\{0,\dots,m\}\rightarrow[0,1]$. We refer to $n \in \{0,\ldots,m\}$ as their \textit{history} and $\sigma(n)$ as their adoption probability at history $n$.

\section{Analysis}\label{sec3}
We first show that, in every equilibrium, uninformed high-cost agents take the risky action only when $n=m$, i.e., 
all of their $m$ immediate predecessors have taken the risky action. We refer to this situation as observing \textit{an $m$-agent herd}, and throughout this section, simply a \textit{herd}. 
\begin{Lemma}\label{Lemma:cutoff1}
Every equilibrium strategy $\sigma$ satisfies $\sigma(n)=0$ for all $n<m$. 
\end{Lemma}
The proof is in Appendix \ref{secA}. Intuitively, $n=m$ is more indicative of the high state than any $n<m$.  Hence, if  uninformed high-cost agents adopted with positive probability at some $n < m$, then they would adopt for sure when $n = m$. Then in the high state, $n=m$ will become absorbing so it will occur almost surely. This implies that observing any $n < m$ would reveal the low state, contradicting uninformed high-cost agents' willingness to adopt there.

 Lemma \ref{Lemma:cutoff1} implies that  uninformed high-cost agents' equilibrium behavior depends only on the comparison between the likelihood ratio of a herd, which measures its informativeness, and 
\begin{equation}\label{R*}
R_* : = \frac{\pi_*/(1-\pi_*)}{\pi_0/(1-\pi_0)} \in (1,\infty).
\end{equation}
Let $L_m^j(x)$ denote
the likelihood ratio of a herd 
in awareness case $j \in \{a,u\}$ when uninformed high-cost agents take action $1$ with probability $x \in [0,1]$ upon observing a herd. 
This gives
\begin{equation}\label{La}
L^{a}_m(x)=\left(\frac{1-(1-p)\gamma}{1-\gamma}\right)^{m}
\cdot\frac{1-(1-p)\,x\,\big[1-(1-\gamma)^m\big]}
{1-x\,\big[1-(1-(1-p)\gamma)^m\big]},
\end{equation}
\begin{equation}\label{Lu}
L^{u}_m(x)=\left(\frac{1-(1-p)\gamma}{1-\gamma}\right)^{m}
\cdot\frac{p(1-p)\,x\,(1-\gamma)^m+\big[1-(1-p)x\big]\big[p+(1-p)\gamma^m\big]}
{p\,x\,(1-(1-p)\gamma)^m+(1-x)\big[p+(1-p)^{m+1}(1-\gamma)\gamma^m\big]}.
\end{equation}
We derive equations (\ref{La}) and (\ref{Lu}) in Appendix \ref{secB}. We now state our main result.
\begin{Theorem}\label{Theorem1}
When $m=1$, $L_1^j(x)$ is constant in $x$ for every $j \in \{a,u\}$. For every $m \geq 2$,
\begin{enumerate}
    \item $L_m^a(x)$ is strictly increasing in $x$.
    \item $L_m^u(x)$ is strictly increasing in $x$ if and only if $\gamma \geq 1/2$ or $\gamma<1/2$ and $p$ is above some threshold $p_m(\gamma) \in (0,1)$. If $\gamma <1/2$ and $p< p_m(\gamma)$, $L_m^u(x)$ is strictly decreasing in $x$.\footnote{If $\gamma<1/2$, then in the knife-edge case where $p=p_m(\gamma)$, $L_m^u(x)$ is constant in $x$.}
\end{enumerate}
\end{Theorem}
The proof is in Appendix \ref{secB}. Together with Lemma \ref{Lemma:cutoff1}, Theorem \ref{Theorem1} leads to a characterization of all equilibria. We state it as Corollary \ref{Cor1}, which follows directly from our theorem. 
\begin{Corollary}\label{Cor1}
In each awareness case $j\in\{a,u\}$, $\sigma(m)=1$ is an equilibrium strategy if and only if $R_*\leq L_m^j(1)$, and  $\sigma(m)=0$ is an equilibrium strategy if and only if $R_*\geq L_m^j(0)$. There is an equilibrium where $\sigma(m) \in (0,1)$ if and only if 
    \begin{enumerate}
        \item either $\min\{L_m^j(0), L_m^j(1)\} < R_* < \max\{L_m^j(0), L_m^j(1)\}$, under which there is exactly one equilibrium that satisfies $\sigma(m) \in (0,1)$, 
        \item or $L_m^j(0) = L_m^j(1) = R_*$, in which case every $\sigma(m) \in [0,1]$ is an equilibrium strategy.
    \end{enumerate}
\end{Corollary}
%
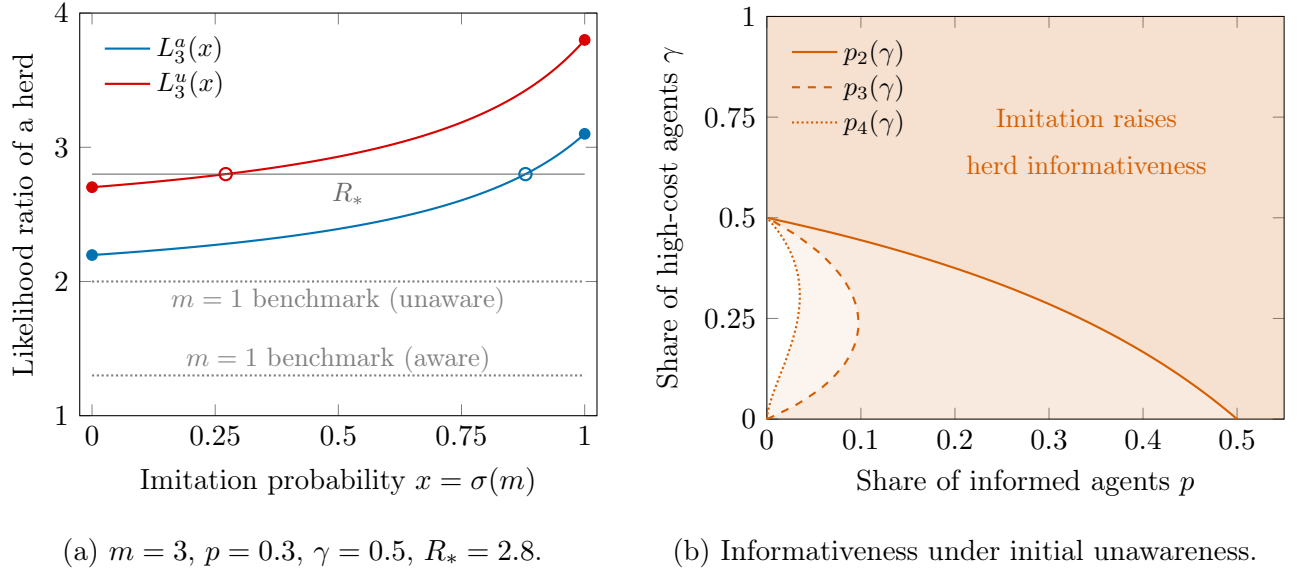
\begin{figure}[t]
\centering
\begin{subfigure}[t]{0.49\textwidth}
\centering
\begin{tikzpicture}
\begin{axis}[
  width=\textwidth, height=0.82\textwidth,
  xlabel={Imitation probability $x = \sigma(m)$},
  ylabel={Likelihood ratio of a herd},
  xmin=-0.025, xmax=1.025, ymin=1, ymax=4,
  xtick={0,0.25,0.5,0.75,1},
  legend pos=north west,
  legend style={draw=none, font=\footnotesize, fill=none},
  legend cell align=left,
]
\addplot[figureAware, thick, domain=0:1, samples=150]
  {LA(x,3,0.3,0.5)};
\addlegendentry{$L_{3}^{a}(x)$}
\addplot[red!85!black, thick, domain=0:1, samples=150]
  {LU(x,3,0.3,0.5)};
\addlegendentry{$L_{3}^{u}(x)$}

\addplot[gray, thick, densely dotted, domain=0:1,
  samples=2, forget plot] {LA(x,1,0.3,0.5)};
\node[gray, font=\footnotesize, anchor=south]
  at (axis cs:0.5,1.24) {$m=1$ benchmark (aware)};

\addplot[gray, thick, densely dotted, domain=0:1,
  samples=2, forget plot] {LU(x,1,0.3,0.5)};
\node[gray, font=\footnotesize, anchor=north]
  at (axis cs:0.5,2.04) {$m=1$ benchmark (unaware)};

\addplot[black!60, thin, domain=0:1, samples=2,
  forget plot] {2.8};
\node[black!60, font=\footnotesize, anchor=north]
  at (axis cs:0.52,2.82) {$R_{*}$};

\addplot[figureAware, only marks, mark=*, mark size=2pt,
  forget plot] coordinates {(0,2.197) (1,3.1)};
\addplot[red!85!black, only marks, mark=*, mark size=2pt,
  forget plot] coordinates {(0,2.7026051) (1,3.8)};
\addplot[figureAware, only marks, mark=o, mark size=2.3pt,
  mark options={fill=white,thick}, forget plot]
  coordinates {(0.8797943,2.8)};
\addplot[red!85!black, only marks, mark=o, mark size=2.3pt,
  mark options={fill=white,thick}, forget plot]
  coordinates {(0.2713423,2.8)};
\end{axis}
\end{tikzpicture}
\caption{$m=3$, $p=0.3$, $\gamma=0.5$, $R_{*}=2.8$.}
\end{subfigure}
\hfill
\begin{subfigure}[t]{0.49\textwidth}
\centering
\begin{tikzpicture}
\def\thresholdTwo{%
    (0.500000,0.000000) (0.493671,0.012500) (0.487179,0.025000) (0.480519,0.037500)
    (0.473684,0.050000) (0.466667,0.062500) (0.459459,0.075000) (0.452055,0.087500)
    (0.444444,0.100000) (0.436620,0.112500) (0.428571,0.125000) (0.420290,0.137500)
    (0.411765,0.150000) (0.402985,0.162500) (0.393939,0.175000) (0.384615,0.187500)
    (0.375000,0.200000) (0.365079,0.212500) (0.354839,0.225000) (0.344262,0.237500)
    (0.333333,0.250000) (0.322034,0.262500) (0.310345,0.275000) (0.298246,0.287500)
    (0.285714,0.300000) (0.272727,0.312500) (0.259259,0.325000) (0.245283,0.337500)
    (0.230769,0.350000) (0.215686,0.362500) (0.200000,0.375000) (0.183673,0.387500)
    (0.166667,0.400000) (0.148936,0.412500) (0.130435,0.425000) (0.111111,0.437500)
    (0.090909,0.450000) (0.069767,0.462500) (0.047619,0.475000) (0.024390,0.487500)
    (0.000000,0.500000)
}
\def\thresholdThree{%
    (0.000000,0.000000) (0.004900,0.005000) (0.018200,0.020000) (0.030100,0.034000)
    (0.040700,0.049000) (0.050100,0.064000) (0.058400,0.078000) (0.065800,0.093000)
    (0.072200,0.108000) (0.077800,0.123000) (0.082700,0.137000) (0.086700,0.152000)
    (0.090100,0.167000) (0.092800,0.181000) (0.094900,0.196000) (0.096300,0.211000)
    (0.097200,0.225000) (0.097400,0.240000) (0.097000,0.255000) (0.096100,0.270000)
    (0.094600,0.284000) (0.092400,0.299000) (0.089800,0.314000) (0.086500,0.328000)
    (0.082600,0.343000) (0.078100,0.358000) (0.073100,0.372000) (0.067300,0.387000)
    (0.061000,0.402000) (0.053900,0.417000) (0.046200,0.431000) (0.037700,0.446000)
    (0.028500,0.461000) (0.018600,0.475000) (0.007800,0.490000) (0.000800,0.499000)
    (0.000000,0.500000)
}
\def\thresholdFour{%
    (0.000000,0.000000) (0.000100,0.005000) (0.001200,0.020000) (0.002700,0.034000)
    (0.004600,0.049000) (0.006700,0.064000) (0.009000,0.078000) (0.011400,0.093000)
    (0.013800,0.108000) (0.016300,0.123000) (0.018700,0.137000) (0.021000,0.152000)
    (0.023200,0.167000) (0.025300,0.181000) (0.027300,0.196000) (0.029100,0.211000)
    (0.030700,0.225000) (0.032100,0.240000) (0.033200,0.255000) (0.034100,0.270000)
    (0.034700,0.284000) (0.035000,0.299000) (0.035000,0.314000) (0.034700,0.328000)
    (0.034000,0.343000) (0.033000,0.358000) (0.031500,0.372000) (0.029700,0.387000)
    (0.027400,0.402000) (0.024700,0.417000) (0.021600,0.431000) (0.017900,0.446000)
    (0.013800,0.461000) (0.009100,0.475000) (0.003900,0.490000) (0.000400,0.499000)
    (0.000000,0.500000)
}
\begin{axis}[
  width=\textwidth, height=0.82\textwidth,
  xlabel={Share of informed agents $p$},
  ylabel={Share of high-cost agents $\gamma$},
  xmin=0, xmax=0.55, ymin=0, ymax=1,
  xtick={0,0.1,0.2,0.3,0.4,0.5},
  ytick={0,0.25,0.5,0.75,1},
  legend pos=north west,
  legend style={draw=none, font=\footnotesize, fill=none},
  legend cell align=left,
]
\addplot[draw=none, fill=figureUnaware, fill opacity=0.065,
  forget plot] coordinates {
  \thresholdTwo (0,1) (0.55,1) (0.55,0)
} \closedcycle;
\addplot[draw=none, fill=figureUnaware, fill opacity=0.065,
  forget plot] coordinates {
  \thresholdThree (0,1) (0.55,1) (0.55,0)
} \closedcycle;
\addplot[draw=none, fill=figureUnaware, fill opacity=0.065,
  forget plot] coordinates {
  \thresholdFour (0,1) (0.55,1) (0.55,0)
} \closedcycle;

\addplot[figureUnaware, thick] coordinates {\thresholdTwo};
\addlegendentry{$p_{2}(\gamma)$}
\addplot[figureUnaware, thick, dashed]
  coordinates {\thresholdThree};
\addlegendentry{$p_{3}(\gamma)$}
\addplot[figureUnaware, thick, densely dotted]
  coordinates {\thresholdFour};
\addlegendentry{$p_{4}(\gamma)$}
\node[font=\footnotesize, text=figureUnaware, align=center]
  at (axis cs:0.34,0.69)
  {Imitation raises\\herd informativeness};
\end{axis}
\end{tikzpicture}
\caption{Informativeness under initial unawareness.}
\end{subfigure}
\caption{Panel (a) illustrates the three equilibria in each awareness case, with filled markers denoting pure and open markers mixed equilibria. Panel (b) shades the region where imitation raises herd informativeness for each displayed $m$, with darker shading indicating overlap.}
\label{fig:theorem1}
\end{figure}
We illustrate Theorem \ref{Theorem1} and Corollary \ref{Cor1} in Figure \ref{fig:theorem1}.
Theorem \ref{Theorem1} implies that, unlike in the case where each agent observes only one predecessor's action, when each agent observes two or more actions (i.e., $m \geq 2$), the informativeness of herds generically varies with the probability that other uninformed agents join them. This in turn makes uninformed agents' actions strategic complements or substitutes, even though there is no payoff externality between agents.

In particular, Theorem \ref{Theorem1} implies that herds can become \textit{strictly more informative} when uninformed agents are more likely to imitate them. This is always the case when agents are automatically aware of action $1$, as well as when they are initially unaware of action $1$ but $\gamma$ or $p$ is large. 
In these situations, uninformed agents' actions are \textit{strategic complements}, 
as each of them benefits more from joining a herd when other uninformed agents join with higher probability. 
Due to this strategic complementarity, there are multiple equilibria when $L_m^j(0) \leq R_* \leq  L_m^j(1)$: 
if $L_m^j(0) < R_* <  L_m^j(1)$, there are two pure-strategy equilibria and a non-trivial mixed-strategy equilibrium, and
if $R_* \in \{L_m^j(0), L_m^j(1)\}$, there are two pure-strategy equilibria.

If agents are initially unaware of action $1$ and $L_m^u(0)>L_m^u(1)$, herds become less informative when uninformed agents join them with higher probability. Uninformed agents' actions are then \textit{strategic substitutes}. 
There is always a unique equilibrium, in which agents play $\sigma(m)=0$ if $R_* \geq L_m^u(0)$, play $\sigma(m)=1$ if $R_* \leq L_m^u(1)$, and mix non-trivially between actions $0$ and $1$ if $R_* \in (L_m^u(1),L_m^u(0))$ with the mixing probability pinned down by agents' indifference condition.

Intuitively, uninformed high-cost agents' imitation can affect the informativeness of herds through three channels. First, it affects their immediate successors' observations and actions. This is the only channel present when $m=1$. Through this channel, more imitation does not affect the informativeness of herds in the steady state, because it raises the adoption probability at a herd by the same amount in both states. To see this, suppose $m=1$. The difference between the adoption probabilities at $n=1$ and at $n=0$ comes entirely from uninformed agents, so this difference, denoted by $\beta$, is independent of the state. The steady-state balancing equation is
\begin{equation*}
\mu_{\theta}(1) = \mu_{\theta}(0) \Pr (a = 1 \mid n = 0, \theta)  + \mu_{\theta} (1) \Big[ \Pr (a = 1 \mid n = 0, \theta) + \beta \Big],
\end{equation*}
which implies that $\mu_{\theta}(1)= \Pr (a = 1 \mid n = 0, \theta) / (1-\beta)$. Hence, $\beta$ does not affect  $\mu_{\overline{\theta}} (1)/ \mu_{\underline{\theta}} (1)$.

Second, when $m \geq 2$, an uninformed agent's imitation also affects whether agents \emph{after} his immediate successor observe a herd. For example, if $m=2$, then unless agent $t$ takes action $1$, agent $t+2$ will not observe a herd regardless of agent $t+1$'s action. Since each agent can only observe the last $m$ actions, the herd he observes might be part of a longer sequence of adoptions. Imitation increases the informativeness of herds because it raises the expected length of such sequences, and longer sequences of adoptions contain more informed choices on average.

Third, when $m \geq 2$ and agents are initially unaware of the risky action, an uninformed agent's imitation also affects whether agents \emph{after} his immediate successor are aware of the risky action. Since uninformed low-cost agents cannot adopt when they are unaware, only informed agents adopt at $n=0$, which makes an adoption at $n=0$ more informative about the state than an adoption at $n \geq 1$. In fact, because of this awareness effect, we have
\begin{equation}\label{comparisonlalu}
L_m^u(x) > L_m^a(x) \textrm{ for every } m \in \mathbb{Z}_{++}, p,\gamma \in (0,1), \textrm{ and } x \in [0,1]. 
\end{equation}
We establish (\ref{comparisonlalu}) in Appendix \ref{secB}. Intuitively, the above inequality says that, holding the parameters and uninformed high-cost agents' strategy fixed, herds are strictly more informative when agents are initially unaware of the risky action than when they are automatically aware of it.

Uninformed high-cost agents' imitation increases the probability that other uninformed agents are aware of action $1$ and, through this channel, lowers the informativeness of herds. This negative effect is more pronounced (i) when most agents have low costs (i.e., $\gamma$ is low), so that a restart at $n=0$ is likely to develop into a herd rather than to terminate midway, and (ii) when most agents are uninformed (i.e., $p$ is low), in which case $n=0$ persists for a long time once it occurs. Uninformed agents' actions are strategic substitutes when the negative effect from raising awareness dominates the positive effect of the second channel.

Next, we compare agents' welfare across equilibria when strategic complementarity in uninformed agents' decisions leads to multiple equilibria. We show that the \textit{highest equilibrium}, the one that maximizes uninformed high-cost agents' probability of imitation, maximizes welfare. 
\begin{Proposition}\label{Prop:welfare2cost}
Fix $j \in \{a,u\}$ and
suppose $L_m^j(0) \leq R_* \leq  L_m^j(1)$. The equilibrium where $\sigma(m)=1$ leads to weakly higher payoffs for all types of agents compared to any other equilibrium, and the uninformed high-cost agents' welfare is strictly higher if and only if $L_m^j(0) \leq R_* <  L_m^j(1)$.
\end{Proposition}
We provide a short proof below. The lesson from this result will be generalized in Section \ref{sec4}.
\begin{proof}
If $L_m^j(0) \leq R_* <  L_m^j(1)$, any uninformed high-cost agent's payoff is $0$ in any equilibrium where $\sigma(m) <1$, since choosing action $0$ is always a best reply. When $R_*<L_m^j(1)$, his payoff is strictly positive in an equilibrium where $\sigma(m)=1$ and is $0$ in other equilibria.
When $R_*=L_m^j(1)$, uninformed high-cost agents' payoff is $0$ in both pure-strategy equilibria. 
When agents are initially unaware of action $1$, uninformed low-cost agents' welfare is also increasing in $\sigma(m)$ 
since an increase in $\sigma(m)$ increases the probability with which they become aware of action $1$. 
\end{proof}

Our next two results examine how the informativeness of herds in the \textit{highest equilibrium}, measured by their likelihood ratio, changes with 
the fraction of high-cost agents $\gamma$, 
the fraction of informed agents $p$, and the number of predecessors' actions each agent can observe $m$. 
\begin{Proposition}\label{Prop:comparativestatics}
Suppose agents are initially aware of action $1$. Herds in the highest equilibrium are strictly more informative as $\gamma$, $p$, or $m$ increases.
\end{Proposition}
\begin{Proposition}\label{Prop:comparativestatics1}
Suppose agents are initially unaware of action $1$. 
\begin{enumerate}
    \item If  $R_* \leq L_m^u(1)$ and $L_m^u(x)$ is strictly increasing in $x$, then herds in the highest equilibrium are strictly more informative as $\gamma$, $p$, or $m$ increases.
    \item If either $R_* > L_m^u(1)$ or $L_m^u(x)$ is strictly decreasing in $x $, then  herds in the highest equilibrium are weakly more informative as $\gamma$ increases.
    But an increase in $p$ or $m$  may strictly decrease the informativeness of herds in the highest equilibrium. 
\end{enumerate}
\end{Proposition}
%
\begin{figure}[t]
\centering
\begin{subfigure}[t]{0.49\textwidth}
\centering
\begin{tikzpicture}
\begin{axis}[
  width=0.96\textwidth, height=0.82\textwidth,
  xlabel={Share of informed agents $p$},
  ylabel={Likelihood ratio of a herd},
  xmin=0, xmax=0.1, ymin=1, ymax=1.45,
  xtick={0,0.02,0.04,0.06,0.08,0.1},
  xticklabel style={/pgf/number format/fixed,
    /pgf/number format/precision=2},
  ytick={1,1.1,1.2,1.3,1.4},
  legend pos=north west,
  legend style={draw=none, font=\footnotesize, fill=none},
  legend cell align=left,
]
\addplot[black!55, thin, domain=0:0.1, samples=2,
  forget plot] {1.3};
\node[black!60, font=\footnotesize, anchor=south east]
  at (axis cs:0.097,1.302) {$R_{*}$};

\addplot[figureAware, thick, domain=0.001:0.1, samples=200]
  {ELL(LA(0,3,x,0.3),LA(1,3,x,0.3),1.3)};
\addlegendentry{Automatically aware}
\addplot[red!85!black, thick, domain=0.001:0.1, samples=400]
  {ELL(LU(0,3,x,0.3),LU(1,3,x,0.3),1.3)};
\addlegendentry{Initially unaware}

\node[red!85!black, font=\scriptsize, anchor=south west]
  at (axis cs:-0.001,1.295) {$x \in (0,1)$};

\node[red!85!black, font=\scriptsize, anchor=north]
  at (axis cs:0.065,1.220) {$x=0$};
\node[figureAware, font=\scriptsize, anchor=north]
  at (axis cs:0.073,1.078) {$x=0$};
\end{axis}
\end{tikzpicture}
\caption{Varying $p$ ($m = 3$).}
\end{subfigure}
\hfill
\begin{subfigure}[t]{0.49\textwidth}
\centering
\begin{tikzpicture}
\begin{axis}[
  width=0.96\textwidth, height=0.82\textwidth,
  xlabel={Number of observed predecessors $m$},
  ylabel={Likelihood ratio of a herd},
  xmin=0.75, xmax=6.25, ymin=1, ymax=1.45,
  xtick={1,2,3,4,5,6},
  ytick={1,1.1,1.2,1.3,1.4},
]
\addplot[black!55, thin, domain=0.75:6.25, samples=2,
  forget plot] {1.3};
\node[black!60, font=\footnotesize, anchor=south east]
  at (axis cs:4.85,1.302) {$R_{*}$};

\addplot[figureAware, thick, mark=*, mark size=1.8pt]
  coordinates {
    (1,1.0214285714) (2,1.0433163265) (3,1.0656731050)
    (4,1.0885089572) (5,1.1118341491)
    (6,1.3749929876)
  };
\addplot[red!85!black, thick]
  coordinates {
    (1,1.4285714286) (2,1.3000000000) (3,1.2328036689)
    (4,1.1547085534) (5,1.1347889276)
    (6,1.3808795655)
  };
\addplot[red!85!black, only marks, mark=*, mark size=1.8pt,
  forget plot]
  coordinates {
    (1,1.4285714286) (3,1.2328036689)
    (4,1.1547085534) (5,1.1347889276)
    (6,1.3808795655)
  };
\addplot[red!85!black, only marks, mark=*, mark size=2.2pt,
  mark options={fill=white,draw=red!85!black,thick}, forget plot]
  coordinates {(2,1.3000000000)};

\node[red!85!black, font=\scriptsize, anchor=west]
  at (axis cs:1.19,1.430) {$x=1$};
\node[red!85!black, font=\scriptsize, anchor=south west]
  at (axis cs:2.0,1.3) {$x \in (0,1)$};
\node[red!85!black, font=\scriptsize, anchor=south]
  at (axis cs:3.75,1.212) {$x=0$};
\node[figureAware, font=\scriptsize, anchor=south]
  at (axis cs:3.7,1.024) {$x=0$};
\node[red!85!black, font=\scriptsize, anchor=south]
  at (axis cs:5.6,1.380) {$x=1$};
\node[figureAware, font=\scriptsize, anchor=south]
  at (axis cs:5.6,1.350) {$x=1$};
\end{axis}
\end{tikzpicture}
\caption{Varying $m$ ($p = 0.05$).}
\end{subfigure}
\caption{Herd informativeness in the highest equilibrium, with $\gamma=0.3$ and $R_{*}=1.3$. Under automatic awareness (i.e., $j=a$), informativeness increases in $p$ (panel (a)) and $m$ (panel (b)); under initial unawareness (i.e., $j=u$), it can decrease in $p$ and $m$ when both $p$ and $m$ are small.}
\label{fig:prop3}
\end{figure}
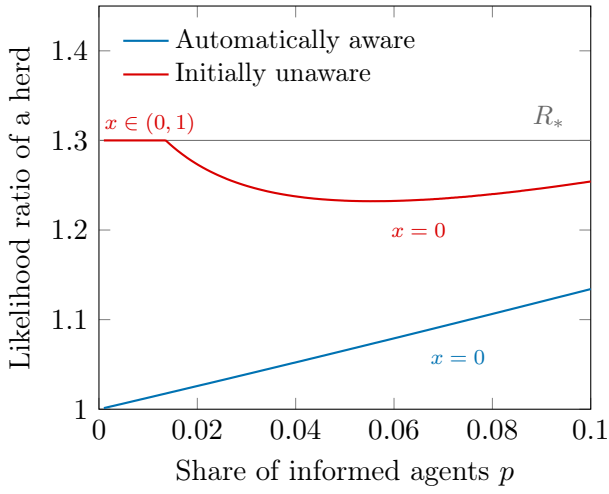
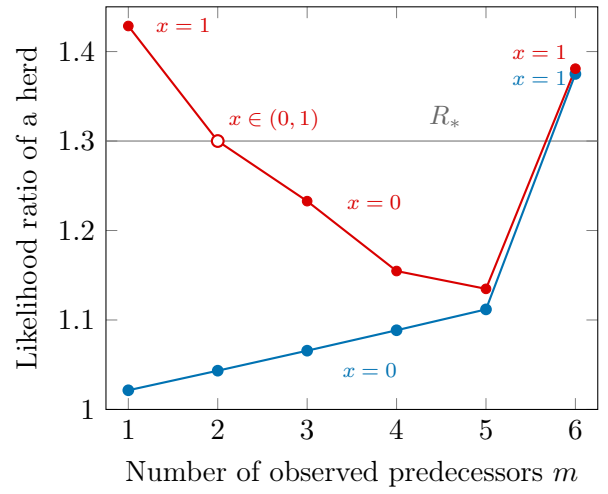
The proofs are in Appendix~\ref{secC}.
The more interesting comparative statics arise when agents are initially unaware of the risky action and either (i) uninformed agents' actions are substitutes, or (ii) their actions are complements but $R_*$ is so large that taking the safe action is always optimal for uninformed high-cost agents. 
Figure \ref{fig:prop3} illustrates how initial unawareness can reverse the comparative statics under automatic awareness. In panel (a), increasing $p$ from $0.01$ to $0.05$ moves the highest equilibrium from partial imitation to no imitation and lowers herd informativeness. In panel (b), increasing $m$ from $3$ to $4$ lowers informativeness even though the highest equilibrium continues to prescribe no imitation.

The countervailing force is that, when agents are initially unaware of the risky action, an increase in $p$ makes $n=0$ less persistent, since adoption will start sooner when more agents are informed, 
and an increase in $m$ makes $n=0$ harder to reach, since it requires $m$ consecutive agents to take the safe action. Both changes reduce the informativeness of herds, because a risky action taken at $n=0$ is more informative about the state than one taken at $n \geq 1$.

This countervailing force dominates when awareness matters more for the informativeness of herds. This is the case, for example, when uninformed agents' actions are substitutes, so that the negative effect of imitation (i.e., preserving awareness) is stronger than its positive effect (i.e., lengthening sequences of adoptions). It is also the case when uninformed high-cost agents have no incentive to take the risky action (i.e., when $R_*$ is large), so that herds are usually short-lived, and their informativeness depends mostly on the restart of adoption at $n=0$.

\section{Agents Who Never Take the Risky Action}\label{sec4}
This section extends the baseline model by introducing agents who never take action $1$. Suppose $c_t$ 
is drawn from $\{\overline{c},c^*,\underline{c}\} \subset \mathbb{R}_+$, with
$\underline{\gamma} \in (0,1)$ the probability that $c_t = \underline{c}$, $\overline{\gamma} \in (0,1)$ 
the probability that $c_t = \overline{c}$, and 
$\gamma := 1-\overline{\gamma}-\underline{\gamma}>0$ the probability that $c_t=c^*$. 
Recall that $\pi_* = \frac{c^*-\underline{\theta}}{\overline{\theta}-\underline{\theta}}$ and $R_* = \frac{\pi_*/(1-\pi_*)}{\pi_0/(1-\pi_0)} $, 
and by (\ref{ass:cost}), $R_* \in (1,\infty)$. 
We assume that $\overline{c} > \overline{\theta}$, so cost-$\overline{c}$ agents never take action $1$.\footnote{When only costs $c^*$ and $\overline{c}$ occur with positive probability, 
the equilibrium is trivial: upon observing any of their last $m$ predecessors 
taking action $1$, uninformed cost-$c^*$ agents learn with certainty that 
the state is $\overline{\theta}$.}
As in the baseline model,
before agent $t \in \mathbb{Z}$ chooses $a_t \in \{0,1\}$ in period $t$, he observes $c_t$, $n_t := \sum_{k=t-m}^{t-1} a_{k} \in \{0,1,...,m\}$, and with probability $p \in (0,1)$, the state $\theta$.  

We focus on the behavior of uninformed cost-$c^*$ agents since others have strictly dominant actions. Let 
$\sigma(n)$ denote the probability that those agents take action $1$ at $n$. We introduce \textit{cutoff strategies}, each of which is 
characterized by a \textit{marginal history}
$k \in \{1, \ldots, m \}$ and a probability $x \in [0,1]$ such that
\begin{equation}\label{eq:cutoff-three}
\sigma_{k,x}(n):=
\begin{cases}
0,&n<k,\\[-6pt]
x,&n=k,\\[-6pt]
1,&n>k.
\end{cases}
\end{equation}
Cutoff strategies are totally ordered by the probability that agents take action $1$. Formally, $\sigma_{k,x}$ is \textit{weakly higher} than $\sigma_{l,y}$ if and only if $k-x \leq l-y$. In this order, 
$\sigma_{m,0}$ is the lowest strategy, $\sigma_{1,1}$ is the highest strategy,\footnote{Since $R_*>1$ and $n=0$ is always a negative signal about the state, uninformed high-cost agents will never take action $1$ when $n=0$. This is why it is without loss of generality to set $\sigma_{1,1}$ to be the highest strategy.} and
strategies $\sigma_{k,0}$ and $\sigma_{k+1,1}$ are equivalent. Lemma \ref{Lemma:cutoff2} shows that focusing on this class of strategies is without loss of generality. 
\begin{Lemma}\label{Lemma:cutoff2}
Every equilibrium strategy is a cutoff strategy. 
\end{Lemma}
The proof is in Appendix \ref{secA}. Unlike in the baseline model, 
uninformed cost-$c^*$ agents may adopt even when some of their last $m$ predecessors took action $0$. 
This is because when $\overline{\gamma}>0$, agents 
will partially attribute the observation $n<m$ to some of their predecessors having cost $\overline{c}$, rather than to unfavorable information about the state.

By Lemma \ref{Lemma:cutoff2}, characterizing equilibria requires comparing,  for each cutoff strategy $\sigma_{k,x}$,
the likelihood ratio of a $k$-agent herd, denoted by $L_{m}^j(k,x)$,
with $R_*$ defined in (\ref{R*}).
To facilitate this comparison,
we then examine
how the informativeness of a $k$-agent herd changes with the probability $x$ with which uninformed cost-$c^*$ agents take action $1$ upon observing it. This dependence also determines whether agents' actions are strategic complements or substitutes.
\begin{Lemma}\label{Lemma:Preliminaries}
Fix any $(m,p,\overline{\gamma},\underline{\gamma})$, $j \in \{a,u\}$, and $k \in \{1, \ldots, m \}$.
\begin{enumerate}
    \item The derivative
    $\partial L_m^j(k,x) / \partial x$ has the same sign for all $x \in [0,1]$. 
    \item If $L_m^j(k,x)$ is weakly increasing in $x$, then
    $L_m^j(k,x) \neq 1$ for all $x \in [0,1]$.
    \item If $k \geq 2$, $L_m^j(k,0) > L_m^j(k-1,0)$. 
\end{enumerate}
\end{Lemma}

The proof is in Appendix \ref{secD}. Hence, for each marginal history $k$, imitation either increases informativeness for all $x \in [0,1]$ or decreases it for all $x$. That is, uninformed cost-$c^*$ agents' actions are either always complements, or always substitutes. When their actions are complements, the likelihood ratio remains on the same side of $1$ for every imitation probability $x \in [0,1]$. Moreover, once we focus on pure cutoff strategies the likelihood ratios are increasing in $k$. 

Theorem \ref{Theorem2} partially extends Theorem \ref{Theorem1}:
\begin{Theorem}\label{Theorem2}
Fix $(p,\overline{\gamma},\underline{\gamma})$.
\begin{enumerate}
    \item If $m=1$, 
   then  for every 
$j \in \{a,u\}$ and in every equilibrium, 
    the marginal history $k$ is $1$ and $L_{1}^j(1,x)$ is constant in $x$.
\item If $m \geq 2$, then for every $j \in \{a,u\}$, there exists $k^j \in \{1,2,...,m\}$  such that $L_{m}^j(k,x)$ is strictly increasing in $x$ 
while $L_m^j(k,0) \geq 1$ if and only if $k>k^j$. Furthermore, $k^a<m$.
\end{enumerate}
\end{Theorem}
The proof is in Appendix \ref{secE}. By Theorem \ref{Theorem2}, when uninformed agents can only observe one predecessor's action, their actions have no impact on the informativeness of herds, which extends our finding in the baseline model.

When agents can observe at least two actions, 
their actions are complements only if the marginal history $k$ exceeds some threshold $k^j$.\footnote{Statement 2 of Theorem \ref{Theorem2} focuses on $k$ such that $L_m^j(k,0) \geq 1$. This is because $R_*>1$, so
in equilibrium, the likelihood ratio of the marginal history must exceed $1$. Furthermore, Lemma \ref{Lemma:Preliminaries} implies that when $\partial L_m^j(k,x) / \partial x>0$, $L_m^j(k,x)$ never equals $1$, which together with the continuity of $L_m^j(k,x)$ with respect to $x$ implies that when $L_m^j(k,0) < 1$, we will have $L_m^j(k,x) < 1$ for every $x \in [0,1]$, i.e., $k$ cannot be the marginal history in equilibrium.} When agents are automatically aware of action $1$, i.e., $j=a$, the threshold $k^a$ is always strictly less than $m$, which implies that uninformed agents' actions are always complements when the marginal history is $m$. This will happen in equilibrium when $R_* \in \big[L_m^a(m,0), L_m^a(m,1)\big]$.
When agents are initially unaware of action $1$, i.e., $j=u$, the threshold $k^u$ might be $m$, in which case 
uninformed agents' actions are strategic substitutes. This is similar to the baseline model where
the informativeness of  $m$-agent herds in the unawareness case can decrease with 
the probability of imitation. When the threshold $k^u$ is less than $m$, uninformed agents' actions are strategic complements under an open set of $R_*$.

Intuitively, when the marginal history $k$ is strictly less than $m$, increasing the probability of imitation has an additional effect on the informativeness of the marginal history, beyond the three effects discussed in the baseline model: it raises the chance that a $k$-agent herd grows into a larger herd, e.g., one that contains $k+1$ risky actions. Since such growth is more likely in the high state, observing a $k$-agent herd rather than one with more adoptions is bad news about the state, and it is worse news when herds with more adoptions are expected to occur with higher probability, i.e., when uninformed agents take the risky action with higher probability.

Theorem \ref{Theorem2} implies that in awareness case $j \in \{a,u\}$, there will be multiple equilibria when 
\begin{equation}\label{multiplicity}
R_* \in \bigcup_{k=k^j+1}^m \Big[
L_m^j(k,0), L_m^j(k,1)
\Big].
\end{equation}
We then derive an analog of Proposition \ref{Prop:welfare2cost}, that when there are multiple equilibria, the one that maximizes uninformed cost-$c^*$ agents' probability of imitation maximizes agents'  welfare. 
\begin{Proposition}\label{Prop:welfare}
Fix any $j \in \{a,u\}$ and $(\overline{\gamma},\underline{\gamma},p,R_*,m)$ such that there are multiple equilibria. 
The highest equilibrium gives uninformed cost-$c^*$ agents a weakly higher payoff relative to other equilibria, and a strictly higher payoff  unless  $\sigma_{m,1}$ is the highest equilibrium 
and $R_*=L_m^j(m,1)$.
\end{Proposition}
The proof is in Appendix \ref{secF}. Proposition \ref{Prop:welfare} implies that the highest equilibrium also maximizes the expected payoff of every type of agent: (i) the expected payoffs of informed agents and of cost-$\overline{c}$ agents do not depend on uninformed cost-$c^*$ agents' strategy, and (ii) the expected payoff of uninformed cost-$\underline{c}$ agents is weakly higher when uninformed cost-$c^*$ agents use a higher strategy, because a higher strategy raises the probability that those agents are aware of action $1$.


Before elaborating on the intuition, we first point out two caveats that distinguish this welfare result from that in the baseline model. First, different equilibria may have different marginal histories,
unlike in the baseline model where the marginal history must be $m$. Second, when the marginal history $k$ is less than $m$, 
an increase in the imitation probability at $k$ may increase the likelihood ratio of $k$ while decreasing those of herds with sizes more than $k$. In fact, for any $i \in \{k+1,\ldots,m\}$, an increase in the imitation probability at $k$ increases the likelihood ratio of $i$ if and only if the rate with which $L_m^j(k,x)$ increases, measured by $\frac{\partial L_m^j (k,x) / \partial x}{ L_m^j (k,x)}$, exceeds 
\begin{equation}\label{increasehighersignals}
\frac{p(1-p)\gamma^2}
{\big(\underline{\gamma}+(1-p)\gamma x\big)
\big(\underline{\gamma}+p\gamma+(1-p)\gamma x\big)}.
\end{equation}
Hence, if $\frac{\partial L_m^j (k,x) / \partial x}{ L_m^j (k,x)}$ is between $0$ and (\ref{increasehighersignals}), 
an increase in agents' imitation probability will benefit those who observe the marginal history while, on average, hurting those who observe larger herds, making it unclear whether a higher equilibrium will result in higher overall welfare.


To understand this welfare result, let us compare uninformed cost-$c^*$ agents' welfare (i) in a \textit{higher equilibrium} where they take action $1$ if and only if $n\geq m-1$, and (ii) in a \textit{lower equilibrium} where they take action $1$ if and only if $n=m$. The general case follows similarly.

In the higher equilibrium, history $m$ occurs with higher probability, although its likelihood ratio may be lower. Nevertheless, because uninformed cost-$c^*$ agents have an incentive to take action $1$ after observing $n=m-1$ in the higher equilibrium, and a herd of size $m$ has a strictly higher likelihood ratio than a herd of size $m-1$, those agents will have a strict incentive to take action $1$ after observing $n=m$. This remains true even though they know that, if others were to play according to the lower equilibrium, they would not have observed 
an $m$-agent herd. Thus, the higher equilibrium always yields weakly higher expected welfare for uninformed cost-$c^*$ agents. This welfare comparison is strict except in the knife-edge case where uninformed cost-$c^*$ agents are indifferent after observing an $m$-agent herd when others follow strategy $\sigma_{m,1}$.

\section{Concluding Remarks}\label{sec5}
This paper studies a social learning model where agents learn from the actions of a limited number of predecessors without knowing the circumstances under which those actions were taken. 
Because agents cannot observe how long a herd has lasted, there can be a \textit{positive} informational externality among uninformed agents: herds become more informative when uninformed agents join those herds with higher probability. This informational externality, in turn, generates strategic complementarity in uninformed agents' strategies, giving rise to multiple steady-state equilibria for a non-degenerate set of parameter values. We conclude this article by discussing some of our modeling assumptions as well as the robustness of our findings.

\paragraph{Observing the Exact Sequence of Actions:} Our baseline model assumes that each agent $t$ can only observe $\sum_{\tau=1}^m a_{t-\tau}$ but not the exact sequence $(a_{t-m},...,a_{t-1})$. We now consider an extension in which agents can observe the exact sequence of the last $m$ actions. 

A first observation is that for every equilibrium in the game where agents cannot observe the sequence of actions, there exists an outcome equivalent equilibrium in the game where agents can observe the exact sequence. This is a consequence of something we show in Appendix \ref{secA}: When each agent $t$'s behavior depends on $(a_{t-m},...,a_{t-1})$ only through $\sum_{\tau=1}^m a_{t-\tau}$, the likelihood ratios of $(a_{t-m},...,a_{t-1})$ and 
 $(a_{t-m}',...,a_{t-1}')$ are the same as long as  $\sum_{\tau=1}^m a_{t-\tau}= \sum_{\tau=1}^m a_{t-\tau}'$.

Next, we show that when agents' costs are drawn from $\{ \underline{c}, c^*\}$, as in the baseline model, regardless of the parameters $(m,p,\gamma,R_*)$ and in both awareness cases, the set of equilibrium outcomes remains unchanged even when agents can observe the exact sequence of the last $m$ actions. This conclusion will be implied by the following proposition:
\begin{Proposition}\label{Prop:order}
For any $m\in\mathbb{Z}_{++}$ and any awareness case $j \in \{a,u\}$, in any equilibrium, uninformed cost-$c^*$ agents will not take action $1$ unless all of their last $m$ predecessors did so. 
\end{Proposition}
The proof is in Appendix \ref{secG}. 
Lastly, when agents' cost is $\overline{c}$ with positive probability, allowing them to observe the exact sequence of the last $m$ actions can lead to additional equilibria. For example, suppose agents are automatically aware of action $1$. Let $m = 4$, $p = 1/2$, $\underline{\gamma} = \overline{\gamma} = 1/4$, $\pi_0 = 1/2$, $\underline{\theta} = 1$, $\overline{\theta} = 2$, $\underline{c} = 1/2$, $c^* = 31/17$, and $\overline{c} = 3$. These parameters give $R_* = 14/3$. The following strategy for uninformed cost-$c^*$ agents is part of an equilibrium:
\[
\sigma(h)=
\begin{cases}
1 & \text{if }h\in \{(0,1,1,1),(1,0,1,1),(1,1,1,0), (1,1,1,1)\},\\
0 & \text{otherwise}.
\end{cases}
\]
It is straightforward to check that $\sigma$ is a steady-state equilibrium. However, $\sigma(0,1,1,1)=1$ whereas $\sigma(1,1,0,1)=0$, even though both histories contain three predecessors who adopted. This equilibrium therefore cannot be represented by a strategy that depends only on the sum of the $m$ predecessors' actions. If agents are not aware of the risky action, the same strategy is an equilibrium if we take $c^*=13/7$, so that $R_*= 6$, and keep all other parameters unchanged. 


\paragraph{Imperfectly Informed Types:} Our baseline model assumes that each informed agent perfectly learns the state. We now 
focus on the case where $c_t \in \{\underline{c},c^*\}$ and assume that 
each agent $t$ is \textit{informed} with probability $p$, and if informed, receives 
a binary private signal $s_t\in\{+,-\}$ with $\Pr(s_t=-\mid\overline\theta) = \Pr(s_t=+\mid\underline\theta)=\varepsilon$ for some $\varepsilon \in (0,1/2)$. Intuitively, signals are ``correct'' with probability $1 - \varepsilon$. 
We assume that agents' private signals are independent of agents' costs, and different informed agents' private signals are independent conditional on $\theta$. 
\begin{Proposition}\label{prop:ImperfectInform}
In awareness case $j \in \{ a, u \}$, for all sufficiently small $\varepsilon > 0$, informed high-cost agents strictly follow their signals in every equilibrium if $R_*<\overline{R}_j$, where 
\begin{equation*}
\overline R_a
 =\frac{(1-p)^{m-1}[p+(1-p)(1-\gamma)^m]}
{(1-\gamma + p \gamma)^m}, \quad
\overline{R}_u
=\left( \frac{1-p}{1-\gamma + p \gamma} \right)^{m-1} \left[ p+(1-p)((1-\gamma)^m+\gamma^m) \right] .
\end{equation*}
\end{Proposition}
The proof is in Appendix \ref{app:ImperfectInform}. Proposition \ref{prop:ImperfectInform} implies that when informed agents receive noisy but sufficiently precise signals, the qualitative features of the equilibria remain unchanged when the required likelihood ratio $R_*$ is not too high. Under our sufficient conditions, one can show that in every equilibrium, informed high-cost agents take action $1$ if and only if their private signal is $+$, and uninformed high-cost agents take action $1$ only if all of their last $m$ predecessors did so. Furthermore, because $R_*>1$, the requirement that $\overline{R}_j > R_*$ also imposes some restrictions on $p$ and $\gamma$. In fact, one can also show that as long as $m \geq 2$ and $\overline{R}_j > R_* >1$, an increase in the probability that uninformed high-cost agents imitate $m$-agent herds strictly increases the informativeness of herds. 

In contrast, when $R_*$ is too high, informed high-cost agents may have an incentive to take action $0$ even upon receiving signal $+$. This can happen, for example, when they observe all of their last $m$ predecessors  took action $0$. 
In fact, a higher $\sigma(m)$ can make $n=0$ stronger evidence that the state is low, reducing informed high-cost agents' incentives to take action $1$. Understanding this additional strategic effect 
is left for future research.

\newpage
\appendix
\section{Proofs of Lemmas \ref{Lemma:cutoff1} and \ref{Lemma:cutoff2}}\label{secA}
We provide a unified proof for Lemmas \ref{Lemma:cutoff1} and \ref{Lemma:cutoff2}. Throughout, $\underline{\gamma} := 1 - \overline{\gamma} - \gamma \geq 0$. The case $\overline{\gamma} = 0$ is the baseline model (Lemma \ref{Lemma:cutoff1}), and $\overline{\gamma} > 0$ is the three-cost extension (Lemma \ref{Lemma:cutoff2}). Let $l(n)$ denote the likelihood ratio of the event that $n$ of the last $m$ agents took action $1$. 
\begin{Lemma}\label{lem:three-stationary}
For any feasible strategy $\sigma : \{ 0, 1, \ldots, m \} \rightarrow [0,1]$, and conditional on any state $\theta$, there exists a unique steady-state distribution $\mu_\theta \in \Delta(\{0,1\}^m)$. The likelihood ratios induced by $\overline{\mu}$ and $\underline{\mu}$ satisfy $l(0)<1<l(m)$ and $l(n)\leq l(n+1)$ for every $n<m$, with strict inequality whenever $l(n+1)>0$. If $\overline{\gamma}>0$, both $\overline{\mu}$ and $\underline{\mu}$ have full support, so $l(0)<l(1)<\ldots<l(m)$.
\end{Lemma}

\begin{proof}[Proof of Lemma \ref{lem:three-stationary}]
Let $H := \{0,1\}^m$ and let $q_\theta(n):=\Pr(a_t=1\mid n_t=n,\theta)$.  When agents are automatically aware of action $1$, $\overline{q}(n) = \underline{\gamma} + \gamma \bigl[p+(1-p)\sigma(n)\bigr]$, and  $\underline{q}(n) = \underline{\gamma}+(1-p)\gamma\sigma(n)$ for every $n \in \{0,1,\ldots,m\}$.  When agents are initially unaware, the above applies for every $n \geq 1$, while $\overline{q}(0) = p(\underline{\gamma} + \gamma) = p(1- \overline{\gamma})$ and $\underline{q}(0) = p\underline{\gamma}$. Since $q_\theta(n)\geq p\underline{\gamma}>0$ for every $n$ and $\theta$, state $\mathbf{1} := (1, 1, \ldots,1) \in H$ can be reached with positive probability in $m$ steps starting from any history, conditional on either state. By Theorem 11.2 in \cite{StokeyLucas1989}, the Markov chain on $H$ has a unique invariant distribution for every feasible strategy $\sigma$ and state $\theta$.

Let $\mu_\theta\in\Delta(H)$ denote the invariant distribution in state $\theta$. Write $\underline{\mu} := \mu_{\underline\theta}$ and $\overline{\mu} := \mu_{\overline\theta}$. For any finite binary sequence $h$, let $|h|$ denote its number of ones and let $\mu_\theta(n) := \sum_{h\in H:\,|h|=n}\mu_\theta(h)$. Since $0 < \underline{q}(n)<1$ for every $n$, any history can be reached from any other by a sequence of $m$ actions in the low state. Hence $\underline{\mu}(h)>0$ for every $h \in H$, and $\underline\mu(n) > 0$ for every $n \leq m$. If $\overline{\gamma} > 0$, then $0 < \overline{q}(n) \leq 1-\overline{\gamma} < 1$ for every $n$, and by the same argument $\overline\mu(n) > 0$ for all $n \leq m$.

We next show that $\mu_\theta(h)$ depends only on $|h|$. For any $n<m$, let $g_\theta(n):=g_\theta(m)\prod_{r=n}^{m-1} \frac{1-q_\theta(r+1)}{q_\theta(r)}$, and choose $g_\theta(m)>0$ so that $\sum_{n=0}^m\binom{m}{n}g_\theta(n)=1$. 
By construction,
\begin{equation}\label{eq:steadystate}
g_{\theta}(n) q_{\theta}(n) = g_{\theta}(n+1) \big[ 1 - q_{\theta}(n+1) \big] \quad \text{for every }n=0,\ldots,m-1.
\end{equation}
If $\overline q(m)=1$, the construction gives
$g_{\overline\theta}(m)=1$ and
$g_{\overline\theta}(n)=0$ for every $n<m$.

We now verify that assigning probability $g_\theta(|h|)$ to each history $h\in H$ gives an invariant distribution. 
Let $0v$ denote the concatenation of action $0$ with the action sequence $v$, and the same applies to $1v$, $v0$, and $v1$. 
For any $v\in\{0,1\}^{m-1}$, the histories $0v$ and $1v$ are the only possible predecessors of $v0$ and $v1$. The proposed distribution assigns these predecessors probabilities $g_\theta(|v|)$ and $g_\theta(|v|+1)$, respectively. Letting $n=|v|$ in \eqref{eq:steadystate} gives $g_\theta(|v|)q_\theta(|v|) = g_\theta(|v|+1)[1-q_\theta(|v|+1)]$. Therefore, the total probability of reaching $v0$ from either $0v$ or $1v$ is
\begin{equation*}
g_\theta(|v|)[1-q_\theta(|v|)] +g_\theta(|v|+1)[1-q_\theta(|v|+1)] =
g_\theta(|v|)[1-q_\theta(|v|)] + g_\theta(|v|)q_\theta(|v|) = g_\theta(|v|).
\end{equation*}
The same identity implies that the total probability of reaching $v1$ from either $0v$ or $1v$ is $g_\theta(|v|+1)$. Every $h \in H$ equals $v0$ or $v1$ for some $v\in\{0,1\}^{m-1}$, so $g_\theta(|h|)$ is invariant. Uniqueness of the invariant distribution therefore implies $\mu_\theta(h)=g_\theta(|h|)$ for every $h \in H$.

Since $\mu_\theta(n)=\binom{m}{n} g_\theta(n)$, our notation is well-defined. Applying \eqref{eq:steadystate} in the two states gives
\begin{equation}\label{eq:lrratio}
l(n) = \frac{\overline{\mu}(n)}{\underline{\mu}(n)}
= \frac{\underline{q}(n)}{\overline{q}(n)} \cdot \frac{1-\overline{q}(n+1)}{1 - \underline{q}(n+1)} l(n+1) \quad \text{for every } n < m.
\end{equation}
All denominators are strictly positive. Since $\overline q(n)=\underline q(n)+p\gamma$ for every $n$, the coefficient multiplying $l(n+1)$ belongs to $[0,1)$. Thus $l(n)\leq l(n+1)$, with strict inequality whenever $l(n+1)>0$. The construction gives $g_{\overline\theta}(m)>0$, so $l(m)>0$ and hence $l(n) \leq l(m-1) < l(m)$ for every $n<m$. If $\overline\gamma>0$, then $l(n+1)>0$ for every $n<m$ since both distributions have full support, which implies that $l(0) < l(1) < \cdots < l(m)$.

Finally, $1=\sum_{n=0}^m \overline{\mu}(n) = \sum_{n=0}^m l(n) \underline{\mu}(n)$. Since $\underline{\mu}(n)>0$ for every $n \leq m$, monotonicity and $l(m-1) < l(m)$ imply that $1$ is strictly between $l(0)$ and $l(m)$, i.e., $l(0)<1<l(m)$.
\end{proof}

Fix an equilibrium strategy $\sigma$. By Lemma \ref{lem:three-stationary}, $l(0) < 1 < R_*$, so $\sigma(0) = 0$ if agents are automatically aware of action $1$. If they are initially unaware, $\sigma(0) = 0$ follows from feasibility.

First suppose $\overline{\gamma} > 0$. If uninformed agents with cost $c^*$ never take action $1$, $\sigma = \sigma_{m,0}$. Otherwise, let $k$ be the smallest $n$ for which $\sigma(n)>0$. Agents' equilibrium incentives imply $l(k) \geq R_* > 0$. By Lemma \ref{lem:three-stationary}, we have $l(n) > l(k)\geq R_*$ for every $n>k$, so $\sigma(n)=1$ at every such $n$. Hence $\sigma = \sigma_{k,x}$ with $x = \sigma(k) \in [0,1]$. This proves Lemma \ref{Lemma:cutoff2}.

Now suppose $\overline{\gamma} = 0$. Assume by way of contradiction that $\sigma(n)>0$ for some $n<m$. Since $\sigma(n)>0$, we must have $l(n) \geq R_*$ as otherwise an uninformed high-cost agent would not take action $1$. Thus $l(m) > l(n) \geq R_*$, so $\sigma(m)=1$. But $\overline{\gamma}=0$ implies that $\overline{q}(m)=1$, so history $\mathbf{1}$ is absorbing in the high state. Uniqueness of the steady-state distribution implies that $\overline{\mu}(m)=1$,  so $l(n)=0<R_*$. This contradicts our earlier conclusion that $l(n) \geq R_*$. This proves Lemma \ref{Lemma:cutoff1}.

\paragraph{Responses to imitation at a cutoff.} For later use, we now derive how the probability of imitation at a cutoff affects the steady-state distribution and the likelihood ratio of the marginal history $k$.  Fix $j \in \{a,u\}$ and $m \geq 1$, and write $L(k,x)$ for $L_m^j(k,x)$. We consider every $k \in \{1,\ldots,m\}$ when $\overline{\gamma} > 0$, and only $k=m$ when $\overline{\gamma} = 0$.

Let $\eta=0$ when agents are aware of action $1$, and $\eta = (1-p) \underline{\gamma}$ when they are initially unaware. Let $q_{\theta,k}(n;x)$ denote the adoption probability after a history of $n$ adoptions under strategy $\sigma_{k,x}$, conditional on state $\theta$, and let $\mu_{\theta,k}(n;x)$ denote the corresponding steady-state probability of a history of $n$ adoptions. When the cutoff $k$ is fixed, we suppress it and write $q_\theta(n;x)$ and $\mu_\theta(n;x)$. 

Under strategy $\sigma_{k,0}$, $\underline{q}(0;0) = \underline{\gamma} - \eta$ and
$\overline{q}(0;0) = \underline{\gamma} + p \gamma - \eta$, while
\[
q_\theta(n;0)=
\begin{cases}
q_\theta(0;0)+\eta, & 1\leq n\leq k, \\
q_\theta(0;0)+\eta+(1-p)\gamma, & k<n\leq m.
\end{cases}
\]
Relative to $\sigma_{k,0}$, the strategy $\sigma_{k,x}$ only changes the adoption probability at the marginal history $k$. Therefore $q_\theta(n;x)=q_\theta(n;0)$ for every $n\neq k$, while $q_\theta(k;x)=q_\theta(k;0)+(1-p)\gamma x$.

Define
\[
C_{\theta,k} =
\frac{\sum_{n=0}^{k-1}\mu_\theta(n;0)}{1-q_\theta(k;0)}
-\frac{\sum_{n=k+1}^{m}\mu_\theta(n;0)}{q_\theta(k;0)}.
\]
Iterating \eqref{eq:steadystate} gives
\begin{equation}\label{eqAppAratioMu}
\frac{\mu_\theta(n;x)}{\mu_\theta(k;x)}
=
\frac{\mu_\theta(n;0)}{\mu_\theta(k;0)}
\begin{cases}
1-\dfrac{(1-p)\gamma x}{1-q_\theta(k;0)}, & n<k,\\[6pt]
1, & n=k,\\[6pt]
1+\dfrac{(1-p)\gamma x}{q_\theta(k;0)}, & n>k.
\end{cases}
\end{equation}
Summing over $n$, using $\sum_{n=0}^m\mu_\theta(n;x)=1$,
and rearranging yields
\begin{equation}\label{eq:D1}
\mu_\theta(k;x)-\mu_\theta(k;0)
=
(1-p)\gamma x\mu_\theta(k;x)C_{\theta,k}.
\end{equation}
Since $(1-p)\gamma x\mu_\theta(k;x)$ is the steady-state probability of imitation by an uninformed cost-$c^*$ agent at the cutoff, $C_{\theta,k}$ measures the net change in the probability of observing $k$ adoptions relative to this probability of added imitation. The reference strategy $\sigma_{k,0}$ already prescribes full imitation above $k$. When $k<m$, additional imitation at $k$ makes histories above $k$ more likely and can reduce the steady-state probability of exactly $k$ adoptions. Thus, $C_{\theta,k}$ can be negative.

Since $\mu_\theta(k;x)>0$ and $\mu_\theta(k;0)>0$,
\eqref{eq:D1} implies $1-(1-p)\gamma xC_{\theta,k} = \mu_\theta(k;0) / \mu_\theta(k;x)>0$, and hence $\mu_\theta(k;x)=\mu_\theta(k;0)/ [1-(1-p)\gamma xC_{\theta,k}]$. Taking the ratio across states gives
\begin{equation}\label{eq:Dlikerat}
L(k,x) = L(k,0)
\frac{1-(1-p)\gamma xC_{\underline\theta,k}}{1-(1-p)\gamma xC_{\overline\theta,k}}.
\end{equation}

For every $n>k$, combining equation \eqref{eqAppAratioMu} with the expression for $\mu_\theta(k;x)$ above gives
\begin{equation}\label{eq:Dupperresponse}
\mu_\theta(n;x)
=
\mu_\theta(n;0)
\frac{1+(1-p)\gamma x/q_\theta(k;0)}
     {1-(1-p)\gamma xC_{\theta,k}}.
\end{equation}
Under strategy $\sigma_{k,0}$, \eqref{eq:steadystate} gives, for every $n>k$,
\begin{equation}\label{eq:Dupperprofile}
\frac{\mu_\theta(n;0)}
     {q_\theta(k;0)\mu_\theta(k;0)}
=
\frac{\binom{m}{n}}{\binom{m}{k}}
\frac{[q_\theta(k;0)+(1-p)\gamma]^{n-k-1}}
     {[1-q_\theta(k;0)-(1-p)\gamma]^{n-k}}.
\end{equation}
The right-hand side is strictly larger in the high state because $\overline q(k;0)>\underline q(k;0)$.

\section{Proof of Theorem \ref{Theorem1}}\label{secB}

We use the notation and identities of Appendix \ref{secA} with $k = m$, so $L_{m}^{j}(x) = L(m,x)$. Recall that under strategy $\sigma_{m,0}$, $q_{\theta}(n; 0) = q_{\theta}(m; 0) = q_{\theta}(0; 0) + \eta$ for every $1 \leq n \leq m$. Because $\overline{\gamma} = 0$ in the baseline model, $\underline{q}(m; 0) = 1 - \gamma$ and $\overline{q}(m; 0) = 1 - \gamma + p \gamma$.

Under $\sigma_{m,0}$, the definition of $g_{\theta}(n)$ in Appendix \ref{secA} and the identity $\mu_{\theta}(n; 0) = \binom{m}{n} g_{\theta}(n)$ give
\begin{align*}
1 &= \sum_{n = 0}^{m} \mu_{\theta}(n; 0) = \mu_{\theta}(m; 0)
\left[ \frac{[1 - q_{\theta}(m; 0)]^{m}}{q_{\theta}(0; 0) q_{\theta}(m; 0)^{m - 1}} 
+ \sum_{n = 1}^{m} \binom{m}{n} \left( \frac{1 - q_{\theta}(m; 0)}{q_{\theta}(m; 0)} \right)^{m - n} \right] \\
&= \frac{\mu_{\theta}(m; 0)}{q_{\theta}(m; 0)^{m}} \left[ 1 + \frac{\eta}{q_{\theta}(0; 0)} [1 - q_{\theta}(m; 0)]^{m} \right],
\end{align*}
where the last equality uses the binomial theorem and $q_{\theta}(m; 0) = q_{\theta}(0; 0) + \eta$. Rearranging yields
\[
\mu_{\theta}(m; 0) =
\frac{q_{\theta}(0; 0) q_{\theta}(m; 0)^{m}}{q_{\theta}(0; 0) + \eta [1 - q_{\theta}(m; 0)]^{m}}.
\]
At $k = m$, the definition of $C_{\theta,k}$ becomes
\begin{equation}\label{eq:Cthetam}
C_{\theta,m} = \frac{1 - \mu_{\theta}(m; 0)}{1 - q_{\theta}(m; 0)}.
\end{equation}
Substituting the preceding expression and rearranging gives
\[
C_{\theta,m} = \sum_{i = 0}^{m - 1} q_{\theta}(m; 0)^{i} 
+ \eta \frac{ q_{\theta}(m; 0)^{m}[1 - q_{\theta}(m; 0)]^{m - 1}}
{q_{\theta}(0; 0) + \eta [1 - q_{\theta}(m; 0)]^{m}}.
\]
Substituting these expressions into \eqref{eq:Dlikerat}
gives \eqref{La} when $\eta = 0$ and \eqref{Lu} when
$\eta = (1 - p)(1 - \gamma)$.

Since $C_{\theta,m}$ does not depend on $x$, differentiating
\eqref{eq:Dlikerat} at $k = m$ yields
\[
\frac{d L_{m}^{j}(x)}{d x}
=
\frac{(1 - p)\gamma L_{m}^{j}(0)
      [C_{\overline{\theta},m} - C_{\underline{\theta},m}]}
     {[1 - (1 - p)\gamma x C_{\overline{\theta},m}]^{2}}.
\]
Thus, the sign of the derivative is the sign of
$C_{\overline{\theta},m} - C_{\underline{\theta},m}$.

When $m = 1$, the expressions above give
$C_{\theta,1} = 1 / (1 - \eta)$, independently of $\theta$.
Hence $L_{1}^{j}(x)$ is constant in $x$ in either awareness case.

Suppose $m \geq 2$ and agents are automatically aware of action $1$. Then $\eta = 0$, so
\[
C_{\overline{\theta},m} - C_{\underline{\theta},m}
=
\sum_{i = 1}^{m - 1}
\big[
(1 - \gamma + p\gamma)^{i} - (1 - \gamma)^{i}
\big]
>
0.
\]
Therefore, $L_{m}^{a}(x)$ is strictly increasing in $x$.

Suppose $m \geq 2$ and agents are initially unaware. Then $\eta = (1 - p)(1 - \gamma)$, so
\begin{align*}
C_{\underline{\theta},m}
&=
\frac{1}{\gamma}
\left[ 1 - \frac{p(1 - \gamma)^{m}}{p + (1 - p)\gamma^{m}} \right], \qquad 
C_{\overline{\theta},m}
= \frac{1}{(1 - p)\gamma} 
\left[ 1 - \frac{p[1 - (1 - p)\gamma]^{m}}{p + (1 - p)^{m + 1}(1 - \gamma)\gamma^{m}}
\right].
\end{align*}
Subtracting and collecting terms gives
\[
C_{\overline{\theta},m} - C_{\underline{\theta},m} =
\frac{p \sum_{i = 0}^{m + 1} g_{i} p^{i}(1 - p)^{m + 1 - i}}{\gamma [p + (1 - p)\gamma^{m}][p + (1 - p)^{m + 1}(1 - \gamma)\gamma^{m}]},
\]
where $g_{0}, \ldots, g_{m+1}$ depend on $m$ and $\gamma$, but not on $p$. Now $g_{0} = \gamma^{m+1}(1 - \gamma)[\gamma^{m-1} - (1 - \gamma)^{m-1}]$ has the sign of $\gamma - 1/2$. We also have $g_{1} = \gamma^{m} [2 - \gamma - m(1 - \gamma)^{m-1} - (1 - \gamma)^{m}]$ and $g_{m+1} = \gamma \sum_{i = 0}^{m-1} [1 - (1 - \gamma)^{i}] > 0$. For $2 \leq i \leq m$, the remaining coefficients satisfy
\begin{align*}
&\frac{g_i}{\binom{m}{i-1}}
= \frac{i-1}{m-i+2}[1-(1-\gamma)^{m-i+2}] -(1-\gamma)^{m-i+1}+(1-\gamma)^m  + \gamma^m[1-\tfrac{m+1-i\gamma}{i}(1-\gamma)^{m-i}]\\
& \geq \gamma^m[1-(1-\gamma)^{m-i+1}] + \frac{m-i+1}{i}(1-\gamma)^{m-i} \left[\frac{i(i-1)}{2}\gamma^2-\gamma^m\right]
\geq \gamma^m[1-(1-\gamma)^{m-i+1}]>0.
\end{align*}
The first inequality follows from $1-(1-\gamma)^{i-1}\leq(i-1)\gamma$, $1-(1-\gamma)^{m-i+2} = \gamma\sum_{\ell=0}^{m-i+1}(1-\gamma)^\ell$ and $\sum_{\ell=0}^{m-i} [(1-\gamma)^\ell-(1-\gamma)^{m-i+1}] \geq (m-i+1)(m-i+2)\gamma(1-\gamma)^{m-i}/2$. The second inequality uses $\gamma^m\leq\gamma^2\leq i(i-1)\gamma^2/2$. Thus, $g_i>0$ for every $2\leq i\leq m+1$.

If $\gamma \geq 1/2$, then $g_{0} \geq 0$. Moreover, $m(1 - \gamma)^{m-1} \leq m 2^{1-m} \leq 1$ gives $g_{1} \geq \gamma^{m} [(1 - \gamma) - (1 - \gamma)^{m}] > 0$.
Hence $C_{\overline{\theta},m} - C_{\underline{\theta},m} > 0$, so $L_{m}^{u}(x)$ is strictly increasing in $x$ for every $p \in (0,1)$.

If $\gamma < 1/2$, then $g_{0} < 0$. In the above expression for $C_{\overline{\theta},m} - C_{\underline{\theta},m}$, everything apart from $\sum_{i = 0}^{m + 1} g_{i} p^{i}(1 - p)^{m + 1 - i}$ is positive. Dividing this sum by $p(1 - p)^{m} > 0$ gives
\[
g_{0} \frac{1 - p}{p} + g_{1} + \sum_{i = 2}^{m + 1} g_{i} \left( \frac{p}{1 - p} \right)^{i - 1}.
\]
This expression is strictly increasing in $p$: the first term is strictly increasing as $g_{0} < 0$, the second is constant, and every term in the sum is strictly increasing because $g_{i} > 0$. This expression tends to $-\infty$ as $p \to 0$ and to $+\infty$ as $p \to 1$. By continuity, the intermediate value theorem gives a zero $p_{m}(\gamma) \in (0,1)$, which is unique by strict monotonicity. Hence $L_{m}^{u}(x)$ is strictly decreasing in $x$ when $p < p_{m}(\gamma)$, strictly increasing when $p > p_{m}(\gamma)$, and constant when $p = p_{m}(\gamma)$.

Finally, when $m=2$, \eqref{eq:Cthetam} gives $C_{\theta,2}=(1+q_{\theta}(0;0)-\eta)/
(1-2\eta+\eta[q_{\theta}(0;0)+\eta])$. Since $\overline{q}(0;0)-\underline{q}(0;0)=p\gamma$, we obtain $C_{\overline{\theta},2}-C_{\underline{\theta},2} = \frac{p\gamma(1-\eta)(1-2\eta)}{[1-\eta(1+\gamma)][1-\eta(1+(1-p)\gamma)]}$. The denominator is positive, so the sign is determined by $1-2\eta$. In the unawareness case, $\eta=(1-p)(1-\gamma)$. Thus, imitation makes a herd more informative precisely when uninformed low-cost agents constitute a minority.

\paragraph{Comparing the two awareness cases.}
We now show that $L_{m}^{u}(x) > L_{m}^{a}(x)$ for every $m \geq 1$, $p,\gamma \in (0,1)$, and $x \in [0,1]$. 
Let $N_a(x)$, $D_a(x)$ and $N_u(x)$, $D_u(x)$ denote the numerator and denominator of the second factor in (\ref{La}) and (\ref{Lu}), respectively. Rearranging gives
\begin{align*}
N_u(x)&=p\,N_a(x)+(1-p)\gamma^m\big[1-(1-p)x\big],\\
D_u(x)&=p\,D_a(x)+(1-p)^{m+1}(1-\gamma)\gamma^m(1-x).
\end{align*}
All four quantities are strictly positive, so $L_m^u(x)>L_m^a(x)$ if and only if $N_u(x)D_a(x)>D_u(x)N_a(x)$. Substituting the two identities and dividing by $(1-p)\gamma^m N_a(x) D_a(x)>0$, this is equivalent to
\begin{equation}\label{awareness}
\frac{1-(1-p)x}{N_a(x)} > (1-p)^m(1-\gamma)\,\frac{1-x}{D_a(x)}.
\end{equation}
For $t\in[0,1]$ and $\alpha\in(0,1)$, let $\phi(t,\alpha):=\frac{1-t}{1-t+t\alpha^m}$. Then $\phi$ is nonnegative, weakly decreasing in $\alpha$, and strictly decreasing in $t$, since $\partial \phi/\partial t=-\alpha^m/(1-t+t\alpha^m)^2<0$. By the definitions of $N_a$ and $D_a$, the left-hand side of (\ref{awareness}) equals $\phi\big((1-p)x,\,1-\gamma\big)$ and the right-hand side equals $(1-p)^m(1-\gamma)\,\phi\big(x,\,1-(1-p)\gamma\big)$. Since $(1-p)x\leq x$ and $1-\gamma<1-(1-p)\gamma$, we have $\phi\big((1-p)x,1-\gamma\big)\geq\phi\big(x,1-(1-p)\gamma\big)\geq 0$, and $\phi\big((1-p)x,1-\gamma\big)>0$ because $(1-p)x<1$. Since $(1-p)^m(1-\gamma)<1$, inequality (\ref{awareness}) follows.

\section{Proofs of Propositions \ref{Prop:comparativestatics}
and \ref{Prop:comparativestatics1}}\label{secC}

By Corollary \ref{Cor1}, the likelihood ratio of a herd in the
highest equilibrium is
\begin{equation}\label{C.1}
\ell\big(L_{m}^{j}(0), L_{m}^{j}(1)\big)
:=
\begin{cases}
L_{m}^{j}(1),
& \text{if } R_{*} \leq L_{m}^{j}(1), \\[-4pt]
\min\{R_{*}, L_{m}^{j}(0)\},
& \text{if } R_{*} > L_{m}^{j}(1).
\end{cases}
\end{equation}
Since $\ell$ is weakly increasing in each argument, it suffices
to examine how $L_{m}^{j}(0)$ and $L_{m}^{j}(1)$ vary with
$\gamma$, $p$, and $m$.
Equations \eqref{La} and \eqref{Lu} give
\begin{alignat*}{2}
L_{m}^{a}(0) &= \left(1 + \frac{p\gamma}{1 - \gamma}\right)^{m},
&\qquad
L_{m}^{a}(1) &= 1 + p\big[(1 - \gamma)^{-m} - 1\big], \\
L_{m}^{u}(0) & = \frac{[1 - (1 - p)\gamma]^{m}}{(1 - \gamma)^{m}} \frac{p + (1 - p)\gamma^{m}}{p + (1 - p)^{m+1}(1 - \gamma)\gamma^{m}},
&\qquad
L_{m}^{u}(1) & = (1 - p) + \frac{p + (1 - p)\gamma^{m}}{(1 - \gamma)^{m}}.
\end{alignat*}

\begin{Lemma}\label{lem:endpoint-comparisons}
\measurelemmahead
\begin{enumerate}[
  label=(\alph*),
  align=left,
  leftmargin=\dimexpr\lemmaheadwidth+ 1em \relax
]
\setlength{\itemindent}{-\lemmaheadwidth}
\item
$L_{m}^{a}(0)$ and $L_{m}^{a}(1)$ are strictly increasing
in $\gamma$, $p$, and $m$.

\setlength{\itemindent}{0pt}
\item
$L_{m}^{u}(0)$ and $L_{m}^{u}(1)$ are strictly increasing
in $\gamma$.

\item
If $m \geq 2$, then $L_{m}^{u}(1)$ is strictly increasing
in $p$.

\item
If $L_{m}^{u}(1) > L_{m}^{u}(0)$, then
$p > (1 - p)\gamma^{m-1}(1 - 2\gamma)$ and
$L_{m+1}^{u}(1) > L_{m}^{u}(1)$.
\end{enumerate}
\end{Lemma}

\begin{proof}
For (a), note that $1 + p \gamma/(1 - \gamma) > 1$ is strictly increasing
in $p$ and $\gamma$, while $(1 - \gamma)^{-m} - 1 > 0$ is strictly increasing in $\gamma$ and $m$. The claims follow from the expressions for $L_{m}^{a}(1)$ and $L_{m}^{a}(0)$.

For (b), $L_{m}^{u}(1)$ is strictly increasing in $\gamma$ because $p + (1 - p)\gamma^{m}$ increases and $(1 - \gamma)^{m}$ decreases.
The first factor in $L_{m}^{u}(0)$ equals $L_{m}^{a}(0)$
and is strictly increasing in $\gamma$ by part (a).
The second factor can be written as
$1 + (1 - p)[1 - (1 - p)^{m}(1 - \gamma)]/
[p\gamma^{-m} + (1 - p)^{m+1}(1 - \gamma)]$.
The added term has a positive numerator that is strictly
increasing in $\gamma$ and a positive denominator that is
strictly decreasing. Thus, $L_{m}^{u}(0)$ is also strictly
increasing in $\gamma$.

For (c), $\frac{\partial L_{m}^{u}(1)}{\partial p} = \frac{1 - (1 - \gamma)^{m} - \gamma^{m}}{(1 - \gamma)^{m}} > 0$ for every $m \geq 2$, as $(1 - \gamma)^{m} + \gamma^{m} < [(1 - \gamma) + \gamma]^{m} = 1$.

For part (d), $L_{m}^{u}(1) > L_{m}^{u}(0)$ implies $m \geq 2$ by Theorem \ref{Theorem1}. Direct computation gives
\begin{equation}\label{eq:Cincrement}
L_{m+1}^{u}(1) - L_{m}^{u}(1) = \frac{p\gamma - (1 - p)\gamma^{m}(1 - 2\gamma)}{(1 - \gamma)^{m+1}},
\end{equation}
which is positive if $\gamma\geq 1/2$.  Consider the case in which $\gamma<1/2$. Assume by way of contradiction that $p\leq (1-p)\gamma^{m-1}(1-2\gamma)$. Then
$p + (1 - p)\gamma^{m}
\leq (1 - p)(1 - \gamma)\gamma^{m-1}$, and hence
\begin{equation}\label{C.2}
p + (1 - p)\gamma^{m} + (1 - p)(1 - \gamma)^{m} \leq
(1 - p)(1 - \gamma)\gamma^{m-1} + (1 - p)(1 - \gamma)^{m}.
\end{equation}
Since $L^u_m(1)>L^u_m(0)$ if and only if
\[
\Delta:=\big[p+(1-p)^{m+1}(1-\gamma)\gamma^m\big]\big[p+(1-p)\gamma^m+(1-p)(1-\gamma)^m\big]-\big(1-(1-p)\gamma\big)^m\big[p+(1-p)\gamma^m\big]>0,
\]
it suffices to show $\Delta<0$. Let
\begin{equation*}
J = p + (1 - p)(1 - \gamma)^{m} - [1 - (1 - p)\gamma]^{m} =
\sum_{i = 2}^{m} \binom{m}{i} \gamma^{i}(1 - \gamma)^{m-i}(p - p^{i}),
\end{equation*}
where the equality follows by expanding $p = p[(1 - \gamma) + \gamma]^{m}$ and $[1 - (1 - p)\gamma]^{m} = [(1 - \gamma) + p\gamma]^{m}$.
In particular, $0 \leq J \leq
p[1 - (1 - \gamma)^{m}
  - m\gamma(1 - \gamma)^{m-1}]$.
Direct substitution gives
\[
\Delta=J\big[p+(1-p)\gamma^m\big]+(1-p)^{m+1}(1-\gamma)\gamma^m\big[p+(1-p)\gamma^m+(1-p)(1-\gamma)^m\big]-(1-p)^2(1-\gamma)^m\gamma^m,
\]
and applying (\ref{C.2}) to the first two terms yields $\Delta\le(1-p)(1-\gamma)\gamma^{m-1}\,(J-K)$, where
\[
K:=(1-p)\gamma\Big[(1-\gamma)^{m-1}\big(1-(1-p)^m(1-\gamma)\big)-(1-p)^m(1-\gamma)\gamma^{m-1}\Big].
\]

We show $J<K$. If $m=2$, then $J=p(1-p)\gamma^2<p(1-p)(1-\gamma)\gamma(2-p)=K$ because $\gamma<1-\gamma$, since $\gamma < 1/2$. If $m \geq 3$, using $1 - (1 - p)^{m}(1 - \gamma) \geq \gamma$ and $(1 - p)^{m} \leq 1$ gives
\[
K \geq (1-p)(1-\gamma)\gamma^2
\big[(1-\gamma)^{m-2}-\gamma^{m-2}\big] \geq
(m-2)(1-p)(1-\gamma)\gamma^{m-1}(1-2\gamma) \geq
p(m-2)(1-\gamma),
\]
where the last inequality uses the contradiction assumption $p\leq(1-p)\gamma^{m-1}(1-2\gamma)$. On the other hand, $J \leq p\left[1 - (1-\gamma)^{m} - m\gamma(1-\gamma)^{m-1}\right]$. For $m = 3$, the bracketed term equals $\gamma^{2}(3-2\gamma) < \gamma < 1-\gamma = (m-2)(1-\gamma)$. For $m \geq 4$, it is less than $1 < (m-2)(1-\gamma)$. Thus, $J < K$, and hence $\Delta < 0$. This contradicts $L_{m}^{u}(1) > L_{m}^{u}(0)$ and proves $p > (1 - p)\gamma^{m-1}(1 - 2\gamma)$. Equation \eqref{eq:Cincrement} then implies $L_{m+1}^{u}(1) > L_{m}^{u}(1)$.
\end{proof}

\paragraph{Proof of Proposition \ref{Prop:comparativestatics}.}
By Lemma \ref{lem:endpoint-comparisons}(a), both $L_{m}^{a}(0)$ and $L_{m}^{a}(1)$ strictly increase when $\gamma$, $p$, or $m$ increases. By Theorem \ref{Theorem1},
$L_{m}^{a}(1) \geq L_{m}^{a}(0)$. Thus, \eqref{C.1} equals $L_{m}^{a}(1)$ if $R_{*} \leq L_{m}^{a}(1)$ and $L_{m}^{a}(0)$ otherwise. The increase is therefore strict in either case. If $L_{m}^{a}(1)$ rises from below $R_{*}$ to at least $R_{*}$, the likelihood ratio increases from $L_{m}^{a}(0) < R_{*}$ to a value of at least $R_{*}$.

\paragraph{Proof of Proposition \ref{Prop:comparativestatics1},
part 1.}
Since $L_{m}^{u}(x)$ is strictly increasing in $x$, Theorem \ref{Theorem1} gives $m \geq 2$ and $L_{m}^{u}(1) > L_{m}^{u}(0)$. As $R_{*} \leq L_{m}^{u}(1)$, Corollary \ref{Cor1} implies that the highest equilibrium is $\sigma_{m,1}$, with likelihood ratio $L_{m}^{u}(1)$. By Lemma \ref{lem:endpoint-comparisons}(b)--(d), $L_{m}^{u}(1)$ strictly increases when $\gamma$, $p$, or $m$ increases; the bound on $p$ in (d) persists at larger $m$, so \eqref{eq:Cincrement} remains positive at every subsequent step. Thus, $R_{*} \leq L_{m}^{u}(1)$ continues to hold, and \eqref{C.1} implies that the likelihood ratio in the highest equilibrium strictly increases.

\paragraph{Proof of Proposition \ref{Prop:comparativestatics1},
part 2.}
By Lemma \ref{lem:endpoint-comparisons}(b),
$L_{m}^{u}(0)$ and $L_{m}^{u}(1)$ strictly increase in
$\gamma$. Since $\ell$ is weakly increasing in each argument,
the likelihood ratio in the highest equilibrium weakly
increases in $\gamma$. For $p$ and $m$, consider the following
examples with $\gamma = 0.3$ and $R_{*} = 1.3$, as in
Figure \ref{fig:prop3}:
\[
\begin{array}{c c c}
(p,m) & L_{m}^{u}(1) & L_{m}^{u}(0) \\ \hline
(0.01,3) & 1.0971 & 1.3214 \\
(0.05,3) & 1.1706 & 1.2328
\end{array}
\qquad
\begin{array}{c c c}
(p,m) & L_{m}^{u}(1) & L_{m}^{u}(0) \\ \hline
(0.05,3) & 1.1706 & 1.2328 \\
(0.05,4) & 1.1903 & 1.1547
\end{array}
\]

In the left table, $L_{m}^{u}(x)$ is strictly decreasing
in $x$. Take $(p,m) = (0.01,3)$, so
$R_{*} = 1.3 \in (L_{3}^{u}(1), L_{3}^{u}(0))$.
After $p$ increases to $0.05$, the likelihood ratio in the
highest equilibrium decreases from $1.3$ to
$L_{3}^{u}(0) \simeq 1.2328$.

In the right table,
$R_{*} = 1.3 > \max\{L_{m}^{u}(0), L_{m}^{u}(1)\}$
in every row. Hence the likelihood ratio in the highest
equilibrium equals $L_{m}^{u}(0)$. It falls from $1.2328$
to $1.1547$ when $m$ increases from $3$ to $4$,
holding $p = 0.05$ fixed.

\section{Proof of Lemma \ref{Lemma:Preliminaries}}\label{secD}

Fix $j \in \{ a, u \}$ and $m \geq 1$, and use the notation and identities of Appendix \ref{secA}, writing $L(k,x)$ for $L_{m}^{j}(k,x)$. Throughout this appendix, $\overline{\gamma} > 0$, so both steady-state distributions have full support by Lemma \ref{lem:three-stationary}. When the cutoff $k$ is fixed, we suppress it in $q_{\theta,k}(n; x)$ and $\mu_{\theta,k}(n; x)$.

For Part 1, differentiating \eqref{eq:Dlikerat} gives
\[
\frac{\partial L(k,x)}{\partial x}
=
\frac{(1 - p)\gamma L(k,0)
      [C_{\overline{\theta},k} - C_{\underline{\theta},k}]}
     {[1 - (1 - p)\gamma x C_{\overline{\theta},k}]^{2}}.
\]
The sign of this derivative is independent of $x$, proving
Part 1.

For Part 2, the definition of $g_{\theta}(n)$ in
Appendix \ref{secA}, the identity
$\mu_{\theta}(n; 0) = \binom{m}{n} g_{\theta}(n)$ under
$\sigma_{k,0}$, and the definition of $C_{\theta,k}$ give
\begin{align*}
\binom{m}{k}
\frac{C_{\theta,k}}{\mu_{\theta}(k; 0)}
&=
\sum_{n = 0}^{k - 1} \binom{m}{n}
\frac{1}{q_{\theta}(n; 0)}
\prod_{i = n + 1}^{k - 1}
\frac{1 - q_{\theta}(i; 0)}{q_{\theta}(i; 0)} -
\sum_{n = k + 1}^{m} \binom{m}{n}
\frac{1}{1 - q_{\theta}(n; 0)}
\prod_{i = k + 1}^{n - 1}
\frac{q_{\theta}(i; 0)}{1 - q_{\theta}(i; 0)}.
\end{align*}
Since $\overline{q}(n; 0) > \underline{q}(n; 0)$ for every $n$,
each term in the first sum is strictly smaller in the high
state, and each term in the second is strictly larger.
The first sum is nonempty, so
\begin{equation}\label{eq:DforAppE}
\frac{C_{\overline{\theta},k}} {\mu_{\overline{\theta}}(k; 0)}
<
\frac{C_{\underline{\theta},k}} {\mu_{\underline{\theta}}(k; 0)}.
\end{equation}
By \eqref{eq:D1},
\begin{align*}
\frac{L(k,x) - 1}{\mu_{\overline{\theta}}(k; x)} 
=  \frac{1}{\mu_{\underline{\theta}}(k; x)} - \frac{1}{\mu_{\overline{\theta}}(k; x)} 
= \frac{1}{\mu_{\underline{\theta}}(k; 0)} - \frac{1}{\mu_{\overline{\theta}}(k; 0)} - (1 - p)\gamma x \left[ \frac{C_{\underline{\theta},k}}{\mu_{\underline{\theta}}(k; 0)} - \frac{C_{\overline{\theta},k}}{\mu_{\overline{\theta}}(k; 0)} \right].
\end{align*}
By \eqref{eq:DforAppE}, the right-hand side is strictly decreasing in $x$. Thus, if $L(k,x_{0}) = 1$ for some $x_{0} \in [0,1]$, then $L(k,x) > 1$ for $x < x_{0}$ and $L(k,x) < 1$ for $x > x_{0}$. Hence $L(k,\cdot)$ cannot be weakly increasing, proving the contrapositive of Part 2.

For Part 3, fix $2 \leq k \leq m$. We compare $L(k-1,0)$ and $L(k,0)$ by relating both to $L(k,1)$. The strategies $\sigma_{k,1}$ and $\sigma_{k-1,0}$ are equivalent. Under this strategy, the adoption probabilities at histories $k-1$ and $k$ are $q_{\theta}(k; 0)$ and $q_{\theta}(k; 1)$, respectively.
Thus, \eqref{eq:lrratio} gives
\[
\frac{L(k,1)}{L(k-1,0)}
= \frac{\overline{q}(k; 0)[1 - \underline{q}(k; 1)]}{\underline{q}(k; 0)[1 - \overline{q}(k; 1)]}.
\]
Furthermore, equation \eqref{eq:Dlikerat} implies
\[
\frac{L(k,1)}{L(k,0)} = \frac{1 - (1 - p)\gamma C_{\underline{\theta},k}}{1 - (1 - p)\gamma C_{\overline{\theta},k}}.
\]
It therefore suffices to show that
\begin{equation}\label{eq:Dsufficient}
\frac{1 - (1 - p)\gamma C_{\underline{\theta},k}}{1 - (1 - p)\gamma C_{\overline{\theta},k}} <
\frac{\overline{q}(k; 0)[1 - \underline{q}(k; 1)]}{\underline{q}(k; 0)[1 - \overline{q}(k; 1)]}.
\end{equation}

Using $\sum_{n = 0}^{m} \mu_{\theta}(n; 0) = 1$, the definition of $C_{\theta,k}$, and $q_{\theta}(k; 1) = q_{\theta}(k; 0) + (1 - p)\gamma$ we get
\begin{align*}
1 - (1 - p)\gamma C_{\theta,k} = \frac{1 - q_{\theta}(k; 1)}{1 - q_{\theta}(k; 0)}  \sum_{n = 0}^{k - 1} \mu_{\theta}(n; 0) + \mu_{\theta}(k; 0)  + \frac{q_{\theta}(k; 1)}{q_{\theta}(k; 0)} \sum_{n = k + 1}^{m} \mu_{\theta}(n; 0).
\end{align*}
The coefficients multiplying $\mu_{\theta}(n; 0)$ are weakly increasing in $n$. By Lemma \ref{lem:three-stationary}, $\overline{\mu}(n; 0)/\underline{\mu}(n; 0)$ is strictly increasing in $n$, hence $\overline{\mu}(\cdot; 0)$ first-order stochastically dominates $\underline{\mu}(\cdot; 0)$. Thus
\begin{align*}
1 - (1 - p)\gamma C_{\underline{\theta},k} \leq \frac{1 - \underline{q}(k; 1)}{1 - \underline{q}(k; 0)} \sum_{n = 0}^{k - 1} \overline{\mu}(n; 0) + \overline{\mu}(k; 0) + \frac{\underline{q}(k; 1)}{\underline{q}(k; 0)} \sum_{n = k + 1}^{m} \overline{\mu}(n; 0).
\end{align*}
Each coefficient in this bound is strictly smaller than $L(k,1)/L(k-1,0)$ times its counterpart in the expression for $1 - (1 - p)\gamma C_{\overline{\theta},k}$, since $\overline{q}(k; x) > \underline{q}(k; x)$. For $n < k$, this is equivalent to $\underline{q}(k; 0)[1 - \overline{q}(k; 0)] < \overline{q}(k; 0)[1 - \underline{q}(k; 0)]$. For $n > k$, it is equivalent to $\underline{q}(k; 1)[1 - \overline{q}(k; 1)] < \overline{q}(k; 1)[1 - \underline{q}(k; 1)]$. At $n = k$, the comparison follows from $L(k,1)/L(k-1,0) > 1$. Consequently,
\[
1 - (1 - p)\gamma C_{\underline{\theta},k} < \frac{L(k,1)}{L(k-1,0)} \big[1 - (1 - p)\gamma C_{\overline{\theta},k}\big].
\]
Dividing by $1 - (1 - p)\gamma C_{\overline{\theta},k} = \mu_{\overline{\theta}}(k; 0)/ \mu_{\overline{\theta}}(k; 1) > 0$ proves \eqref{eq:Dsufficient} and hence Part 3.

The same argument applies when $k = 1$, replacing $L(k,1)/L(k-1,0)$ in the coefficient comparisons by $\overline{q}(1; 0)[1 - \underline{q}(1; 1)]/ \{\underline{q}(1; 0)[1 - \overline{q}(1; 1)]\}$. Thus, \eqref{eq:Dsufficient} holds for every $k \in \{1, \ldots, m\}$.

For $k < m$, the strategies $\sigma_{k,0}$ and $\sigma_{k+1,1}$ coincide. Moreover, the adoption probability at the cutoff does not depend on which cutoff is used: $q_{\theta,k}(k; x) = q_{\theta,k+1}(k+1; x)$ for every $x \in [0,1]$. Thus, \eqref{eq:lrratio}, \eqref{eq:Dlikerat}, and \eqref{eq:Dsufficient} imply
\begin{equation}\label{eq:Dendpointorder}
\frac{L(k+1,1)}{L(k,0)}
=
\frac{q_{\overline{\theta},k}(k; 0)[1 - q_{\underline{\theta},k}(k; 1)]}{q_{\underline{\theta},k}(k; 0)[1 - q_{\overline{\theta},k}(k; 1)]}
> \max\left\{1, \frac{L(k,1)}{L(k,0)}\right\}.
\end{equation}
In particular, $L(k,1)$ is strictly increasing in $k$.

\section{Proof of Theorem \ref{Theorem2}}\label{secE}

We use the notation and identities of Appendices \ref{secA} and \ref{secD}. Recall that $q_{\theta}(k; 0) = q_{\theta}(0; 0) + \eta$.

Suppose that $m = 1$. Since $l(0) < 1 < R_{*}$, uninformed cost-$c^{*}$ agents choose action $0$ after observing $n = 0$, so every equilibrium is represented by $\sigma_{1,x}$. Appendix \ref{secB} gives $C_{\theta,1} = 1/(1 - \eta)$ for all $\theta$, and this computation does not require $\overline{\gamma} = 0$. By equation \eqref{eq:Dlikerat},
$L_{1}^{j}(1, x)$ is constant in $x$. This proves Part 1.

For Part 2, suppose that $m \geq 2$. Let $\mathcal{K}$
denote the set of $k \in \{1, \ldots, m\}$ such that
$L(k, \cdot)$ is strictly increasing and $L(k, 0) \geq 1$.

\begin{Lemma}\label{lem:K}
For any $k \in \mathcal{K}$ and $x \in [0,1]$, we have $L(k, x) > 1$, $0 < C_{\underline{\theta},k} < C_{\overline{\theta},k}$, and 
\[
0 < \frac{C_{\overline{\theta},k}}{\mu_{\overline{\theta}}(k; 0)} < \frac{C_{\underline{\theta},k}}{\mu_{\underline{\theta}}(k; 0)}.
\]
\end{Lemma}

\begin{proof}
Fix $k \in \mathcal{K}$. By definition, $L(k, \cdot)$ is strictly increasing and $L(k, 0) \geq 1$. Lemma \ref{Lemma:Preliminaries}(2) rules out $L(k, 0) = 1$. Therefore, $L(k, x) \geq L(k, 0) > 1$ for every $x \in [0,1]$.

By definition of $\mathcal{K}$, $L(k, 1) > L(k, 0)$. Together with \eqref{eq:Dlikerat} at $x = 1$, this implies $C_{\underline{\theta},k} < C_{\overline{\theta},k}$. Equation \eqref{eq:DforAppE} gives $C_{\overline{\theta},k}/\mu_{\overline{\theta}}(k; 0) < C_{\underline{\theta},k}/\mu_{\underline{\theta}}(k; 0)$, or equivalently
$C_{\overline{\theta},k} < L(k, 0) C_{\underline{\theta},k}$.
Hence $C_{\underline{\theta},k} < C_{\overline{\theta},k} < L(k, 0) C_{\underline{\theta},k}$. Since $L(k, 0) > 1$, this implies $C_{\underline{\theta},k} > 0$, proving both inequalities.
\end{proof}

It suffices to show that (i) $1 \notin \mathcal{K}$; (ii) $k \in \mathcal{K}$ and $k < m$ imply $k + 1 \in \mathcal{K}$; (iii) $m \in \mathcal{K}$ if $j = a$.

To prove (i), assume by way of contradiction that $1 \in \mathcal{K}$. By Lemma \ref{lem:K}, $C_{\overline{\theta},1} > C_{\underline{\theta},1} > 0$. Equation \eqref{eq:Dupperprofile} implies that $\sum_{n = 2}^{m} \mu_{\theta}(n; 0)/ [q_{\theta}(1; 0)\mu_{\theta}(1; 0)]$ is strictly larger in the high state.

By definition,
$C_{\theta,1}
= \mu_{\theta}(0; 0)/[1 - q_{\theta}(1; 0)]
- \sum_{n = 2}^{m} \mu_{\theta}(n; 0)/q_{\theta}(1; 0)$.
Using \eqref{eq:steadystate} at $n = 0$ and
$\mu_{\theta}(n; 0) = \binom{m}{n} g_{\theta}(n)$ yields
$C_{\theta,1}
= \mu_{\theta}(1; 0)/[m q_{\theta}(0; 0)]
- \sum_{n = 2}^{m} \mu_{\theta}(n; 0)/q_{\theta}(1; 0)$.
Because $\mu_{\theta}(\cdot; 0)$ is a probability distribution, $1 = \sum_{n = 0}^{m} \mu_{\theta}(n; 0) = \mu_{\theta}(1; 0)
  \left[1 + \frac{1}{m q_{\theta}(0; 0)}\right]
  - q_{\theta}(1; 0) C_{\theta,1}$. Combining this with the preceding expression for
$C_{\theta,1}$ yields
\[
\frac{\sum_{n = 2}^{m} \mu_{\theta}(n; 0)}
     {q_{\theta}(1; 0)\mu_{\theta}(1; 0)}
=
\frac{
1 - [1 - \eta + (m - 1)q_{\theta}(0; 0)]C_{\theta,1}
}{
m q_{\theta}(0; 0)
[1 + q_{\theta}(1; 0)C_{\theta,1}]
}.
\]
By Lemma \ref{lem:K}, $C_{\theta,1} > 0$, so the denominator is positive in every state. The numerators are also positive because the left-hand side is positive. Since
$C_{\overline{\theta},1} > C_{\underline{\theta},1} > 0$ and both $q_{\theta}(0; 0)$ and $q_{\theta}(1; 0)$ are larger in the high state, the right-hand side is strictly smaller in the high state. This contradicts \eqref{eq:Dupperprofile}, proving statement (i).

For statement (ii), fix $k \in \mathcal{K}$ with $k < m$. We now write the cutoff index in $q_{\theta,k}(n; x)$ and $\mu_{\theta,k}(n; x)$. Since $q_{\theta,k}(0; 0)$ does not depend on $k$, we continue to write it as $q_{\theta}(0; 0)$. Thus, $q_{\theta,k}(k; 0) = q_{\theta,k+1}(k + 1; 0) = q_{\theta}(0; 0) + \eta$. Applying the expression for $\binom{m}{k}C_{\theta,k}/\mu_{\theta,k}(k; 0)$ derived in Appendix \ref{secD} at cutoffs $k$ and $k + 1$, and rearranging, gives
\begin{equation}\label{eq:E1}
\begin{aligned}
\frac{q_{\theta,k}(k; 0) C_{\theta,k+1}}
     {\mu_{\theta,k+1}(k + 1; 0)}
&=
\frac{k + 1}{m - k}
[1 - q_{\theta,k}(k; 0)]
\frac{C_{\theta,k}}{\mu_{\theta,k}(k; 0)}
+ \frac{k + 1}{m - k}
+ \frac{1 - q_{\theta,k}(k; 0)}
       {1 - q_{\theta,k}(k; 1)}
\\
&\quad+
\frac{(1 - p)\gamma}{\binom{m}{k + 1}}
\sum_{n = k + 2}^{m} \binom{m}{n}
\frac{q_{\theta,k}(k; 1)^{n - k - 2}}
     {[1 - q_{\theta,k}(k; 1)]^{n - k}}.
\end{aligned}
\end{equation}
By Lemma \ref{lem:K}, $C_{\theta,k} > 0$, so the
right-hand side is strictly positive and
$C_{\theta,k+1} > 0$ for every $\theta$.

Divide \eqref{eq:E1} by $[1 - q_{\theta,k}(k; 1)]\,C_{\theta,k}/\mu_{\theta,k}(k; 0) > 0$, which is strictly smaller in the high state by \eqref{eq:DforAppE} and $q_{\overline{\theta},k}(k; 1) > q_{\underline{\theta},k}(k; 1)$. The first term on the right-hand side is $\frac{k + 1}{m - k}\,\frac{1 - q_{\theta,k}(k; 0)}{1 - q_{\theta,k}(k; 1)}$. Its second factor equals $1 + (1 - p)\gamma/[1 - q_{\theta,k}(k; 1)]$, which is strictly larger in the high state. The other terms of \eqref{eq:E1} are positive and weakly larger in the high state: $(k + 1)/(m - k)$ is constant, $[1 - q_{\theta,k}(k; 0)]/[1 - q_{\theta,k}(k; 1)]$ is strictly larger as just shown, and each term in the sum is strictly increasing in $q_{\theta,k}(k; 1)$. Dividing them by the strictly smaller divisor makes each strictly larger in the high state as well.
So,
\[
\frac{q_{\overline{\theta},k}(k; 0)}{1 - q_{\overline{\theta},k}(k; 1)}
\frac{ C_{\overline{\theta},k+1}/ \mu_{\overline{\theta},k+1}(k + 1; 0)}
{ C_{\overline{\theta},k}/ \mu_{\overline{\theta},k}(k; 0) }
>
\frac{q_{\underline{\theta},k}(k; 0)}{1 - q_{\underline{\theta},k}(k; 1)}
\frac{ C_{\underline{\theta},k+1}/ \mu_{\underline{\theta},k+1}(k + 1; 0) }
{ C_{\underline{\theta},k}/ \mu_{\underline{\theta},k}(k; 0) }.
\]
Combining this with \eqref{eq:Dendpointorder} yields
\[
L(k + 1, 1) \frac{ C_{\overline{\theta},k+1}/ \mu_{\overline{\theta},k+1}(k + 1; 0) }
{ C_{\underline{\theta},k+1}/ \mu_{\underline{\theta},k+1}(k + 1; 0) }
> L(k, 0) \frac{ C_{\overline{\theta},k}/ \mu_{\overline{\theta},k}(k; 0) }
{ C_{\underline{\theta},k}/ \mu_{\underline{\theta},k}(k; 0) }
= \frac{C_{\overline{\theta},k}}{C_{\underline{\theta},k}} > 1.
\]
In particular, the left-hand side is greater than $1$. Using \eqref{eq:Dlikerat}, $L(k + 1, 0) = \mu_{\overline{\theta},k+1}(k + 1; 0)/ \mu_{\underline{\theta},k+1}(k + 1; 0)$, and $1 - (1 - p)\gamma C_{\theta,k+1} > 0$ by \eqref{eq:D1}, this inequality simplifies to $C_{\overline{\theta},k+1} > C_{\underline{\theta},k+1}$. By \eqref{eq:Dlikerat}, $L(k + 1, \cdot)$ is strictly increasing, and Lemma \ref{Lemma:Preliminaries}(3) gives $L(k + 1, 0) > L(k, 0) \geq 1$. Hence $k + 1 \in \mathcal{K}$, proving statement (ii).

For statement (iii), let $j = a$, so $\eta = 0$. The calculation in Appendix \ref{secB} under $\sigma_{m,0}$ gives $\mu_{\theta,m}(m; 0) = q_{\theta}(0; 0)^{m}$ and $C_{\theta,m} = \sum_{i = 0}^{m - 1} q_{\theta}(0; 0)^{i}$. Hence $L(m, 0) = [\overline{q}(0; 0)/\underline{q}(0; 0)]^{m} > 1$. Moreover, $C_{\theta,m}$ is strictly increasing in $q_{\theta}(0; 0)$ since $m \geq 2$. Thus $C_{\overline{\theta},m} > C_{\underline{\theta},m}$, and \eqref{eq:Dlikerat} implies that $L(m, x)$ is strictly increasing in $x$. Hence $m \in \mathcal{K}$, proving statement (iii).

Finally, define $k^{j} = m$ if $\mathcal{K} = \emptyset$ and $k^{j} = \min \mathcal{K} - 1$ otherwise. Statements (i)--(iii) give $k^{j} \in \{1, \ldots, m\}$, $\mathcal{K} = \{k \in \{1, \ldots, m\}: k > k^{j}\}$, and $k^{a} < m$, proving Part 2 of the theorem.

\section{Proof of Proposition \ref{Prop:welfare}}\label{secF}

Fix $j \in \{a,u\}$ and $m \geq 1$. Since $\overline{\gamma} > 0$, both steady-state distributions have full support and the likelihood ratio is strictly increasing in $n$ by Lemma \ref{lem:three-stationary}. Part 3 of Lemma \ref{Lemma:Preliminaries}  and \eqref{eq:Dendpointorder} also imply that both $L(k, 0)$ and $L(k, 1)$ are strictly increasing in $k$.

For any strategy $\sigma$, let $\overline{\mu}_{\sigma}$ and $\underline{\mu}_{\sigma}$ denote its steady-state distributions in the high and low states, respectively, and write $z_{\sigma}(n) := \overline{\mu}_{\sigma}(n) - R_{*}\underline{\mu}_{\sigma}(n)$.
By the definition of $R_{*}$, the expected payoff of an uninformed cost-$c^{*}$ agent is $\Pi(\sigma) = \pi_{0}(\overline{\theta} - c^{*}) \sum_{n = 0}^{m} \sigma(n)z_{\sigma}(n)$. In equilibrium, $z_{\sigma}(n) \geq 0$ whenever $\sigma(n) > 0$, with equality whenever $0 < \sigma(n) < 1$.

\paragraph{Ordering the equilibria.}
The equilibrium set is compact and, by Lemma \ref{Lemma:cutoff2}, totally ordered, so it has a highest element $\sigma^{H} = \sigma_{k,x}$. Any lower equilibrium $\sigma' = \sigma_{k',y}$, with $x > 0$ and $y < 1$, has $k' \geq k$, with $y < x$ if $k' = k$. Agents' equilibrium incentives imply $L(k, x) \geq R_{*} \geq L(k', y)$.

We claim that $x = 1$. Assume by way of contradiction that $0 < x < 1$, so $L(k, x) = R_{*}$. If $L(k, \cdot)$ is weakly increasing, then $L(k, 1) \geq R_{*}$.
When $k \geq 2$, Lemma \ref{Lemma:Preliminaries} part 3 also gives $L(k - 1, 0) < L(k, 0) \leq R_{*}$. When $k = 1$, action $0$ at $n = 0$ is optimal because $l(0) < 1 < R_{*}$. Thus $\sigma_{k,1}$ is an equilibrium, contradicting that $\sigma_{k,x}$ is the highest equilibrium.

If, instead, $L(k, \cdot)$ is strictly decreasing, then $L(k, 0) > L(k, x) = R_{*}$. If $k' = k$, then $y < x$ gives $L(k', y) > R_{*}$. If $k' > k$, Lemma \ref{Lemma:Preliminaries} part 3 and \eqref{eq:Dendpointorder} imply $L(k', 0) > L(k, 0) > R_{*}$ and $L(k', 1) > L(k' - 1, 0) \geq L(k, 0) > R_{*}$. Monotonicity therefore gives $L(k', y) > R_{*}$ in this case as well, contradicting agents' incentives to choose action $0$ with positive probability at $k'$ under $\sigma'$. Hence $\sigma^{H} = \sigma_{k,1}$.

Moreover, $L(k', 1) \geq L(k, 1) \geq R_{*} \geq L(k', y)$. Since $y < 1$, $L(k', \cdot)$ cannot be strictly decreasing, so $L(k', 0) \leq L(k', y) \leq R_{*}$.
As both $L(i, 0)$ and $L(i, 1)$ are increasing in $i$, we have
\begin{equation}\label{eq:Finterval}
L(i, 0) \leq R_{*} \leq L(i, 1) \qquad \text{for every } i \in \{k, \ldots, k'\}.
\end{equation}

\paragraph{Payoffs above the cutoff.}
For $i < m$ satisfying \eqref{eq:Finterval}, we show that,
for every $x \in [0,1)$,
\begin{equation}\label{eq:Fgain}
z_{\sigma_{i,1}}(n) > z_{\sigma_{i,x}}(n)
\qquad \text{for every } n > i.
\end{equation}

By \eqref{eq:Finterval} and Lemma \ref{Lemma:Preliminaries}(1), $L(i, \cdot)$ is nondecreasing. Since $L(i, 1) \geq R_{*} > 1$, continuity of $L(i, \cdot)$ and Lemma \ref{Lemma:Preliminaries} part 2 imply $L(i, 0) > 1$. 
Equations \eqref{eq:Dlikerat} and \eqref{eq:DforAppE} give $C_{\underline{\theta},i} \leq C_{\overline{\theta},i} < L(i, 0) C_{\underline{\theta},i}$. These inequalities imply
\begin{equation}\label{eq:Fpositive}
0 < C_{\underline{\theta},i} \leq C_{\overline{\theta},i}.
\end{equation}

To establish \eqref{eq:Fgain}, we compare across states the increase in the probability of observing each $n > i$ when imitation at the cutoff $i$ rises from $x$ to $1$. For $n > i$, define
$\Delta_{\theta}(n; x) = \mu_{\theta,i}(n; 1) - \mu_{\theta,i}(n; x)$. Subtracting \eqref{eq:Dupperresponse} at $x$ from \eqref{eq:Dupperresponse} at $1$ gives
\begin{equation}\label{eq:Fprobabilitygain}
\Delta_{\theta}(n; x) =
\frac{ (1 - p)\gamma(1 - x)\mu_{\theta,i}(n; 0)[1 + q_{\theta,i}(i; 0)C_{\theta,i}]}
{ q_{\theta,i}(i; 0) [1 - (1 - p)\gamma C_{\theta,i}] [1 - (1 - p)\gamma x C_{\theta,i}] } > 0.
\end{equation}
The denominator is strictly positive by \eqref{eq:D1} for every $\theta$. Taking the ratio and using \eqref{eq:Dlikerat} yields
\begin{equation}\label{eq:Fgainratio}
\frac{\Delta_{\overline{\theta}}(n; x)}{\Delta_{\underline{\theta}}(n; x)}
= \frac{ \mu_{\overline{\theta},i}(n; 0)/ [q_{\overline{\theta},i}(i; 0) \mu_{\overline{\theta},i}(i; 0)] } { \mu_{\underline{\theta},i}(n; 0)/ [q_{\underline{\theta},i}(i; 0) \mu_{\underline{\theta},i}(i; 0)] }
\times
\frac{ 1 + q_{\overline{\theta},i}(i; 0)C_{\overline{\theta},i}}{ 1 + q_{\underline{\theta},i}(i; 0)C_{\underline{\theta},i} }
\times
\frac{L(i, x)}{L(i, 0)}L(i, 1) > R_{*}.
\end{equation}
The first factor exceeds $1$ by \eqref{eq:Dupperprofile}. The second exceeds $1$ as $q_{\overline{\theta},i}(i; 0) > q_{\underline{\theta},i}(i; 0)$ and \eqref{eq:Fpositive} holds.
Moreover, $L(i, x)/L(i, 0) \geq 1$ and $L(i, 1) \geq R_{*}$. So, $z_{\sigma_{i,1}}(n) - z_{\sigma_{i,x}}(n) = \Delta_{\overline{\theta}}(n; x) - R_{*}\Delta_{\underline{\theta}}(n; x) > 0$, proving \eqref{eq:Fgain}. Thus full imitation at $i$ raises the payoff contribution of every history $n > i$.

\paragraph{Comparing equilibrium payoffs.}
Since $y < 1$, either $y = 0$ or agents are indifferent at the marginal history $k'$ in $\sigma'$. Thus $\Pi(\sigma') = \pi_{0}(\overline{\theta} - c^{*}) \sum_{n = k' + 1}^{m} z_{\sigma'}(n)$. In the highest equilibrium, all histories at which agents adopt give nonnegative payoff, so $\Pi(\sigma^{H}) \geq \pi_{0}(\overline{\theta} - c^{*}) \sum_{n = k' + 1}^{m} z_{\sigma^{H}}(n)$.

If $k'<m$, start from $\sigma'=\sigma_{k',y}$ and compare it to $\sigma_{k',1}$. If $k'>k$, observe that $\sigma_{k',1} = \sigma_{k'-1,0}$ and compare it to $\sigma_{k'-1,1}$. Repeat until reaching $\sigma^H=\sigma_{k,1}$. Equation \eqref{eq:Finterval} ensures that \eqref{eq:Fgain} applies at each step. Since every $n>k'$ lies above every cutoff encountered, each step strictly increases its payoff contribution. Hence $z_{\sigma^H}(n)>z_{\sigma'}(n)$ for every $n>k'$. Since the sum over $n>k'$ is nonempty, $\Pi(\sigma^H)>\Pi(\sigma')$.

If $k' = m$, $\Pi(\sigma') = 0$ by definition. Strict monotonicity of the likelihood ratios implies $\Pi(\sigma^{H}) > 0$ unless $k = m$ and $L(m, 1) = R_{*}$, in which case every equilibrium gives zero payoff. This proves the proposition.

\section{Proof of Proposition \ref{Prop:order}}\label{secG}
We establish the proposition using two lemmas. We let $\mathbf{0} = (0,0,\ldots,0)$ and $\mathbf{1} = (1,1,\ldots,1)$.
Define $\underline{\alpha}=1-\gamma$, 
$\overline{\alpha}=1-(1-p)\gamma$, and $\delta=(1-p)\gamma$. Since $p,\gamma\in(0,1)$, we have $0<\underline{\alpha} < \overline{\alpha} < 1$. Let $q_{\theta}(h)$ be the probability of action $1$ after history $h$, conditional on state $\theta$. When agents are automatically aware, for every $h \in H$
\begin{equation}\label{eq:order_transition}
q_{\theta}(h)=\alpha_\theta+\delta\sigma(h).
\end{equation}
When agents are initially unaware, \eqref{eq:order_transition} holds for $h \neq \mathbf{0}$, while $\underline{q}(\mathbf{0}) = p(1-\gamma)$, $\overline{q}(\mathbf{0}) = p$, and $\sigma(\mathbf{0})=0$. For any feasible strategy and its steady-state distributions, write $z(h)=\overline{\mu}(h)-R_*\underline{\mu}(h)$. In equilibrium, at every history apart from $\mathbf{0}$, agents behave optimally if and only if
\begin{equation}\label{eq:order_optimality}
\sigma(h)z(h)\ge0,
\qquad
[1-\sigma(h)]z(h)\le0.
\end{equation}
In particular, at such histories, $\sigma(h)>0$ implies $z(h)\ge0$, and $z(h)>0$ implies $\sigma(h)=1$.

\begin{Lemma}\label{lem:order_steady_states}
For any feasible strategy, $\underline{\mu}$ and $\overline{\mu}$ are unique, with $\underline{\mu}(h)>0$ for every $h\in H$. In every equilibrium, $\sigma(h)=0$ at every history ending in zero.
\end{Lemma}
\begin{proof}
In either awareness case, $0<\underline{q}(h)<1$ for every $h$. Any ordered history can therefore be reached from any other by a sequence of $m$ actions in the low state. The low-state Markov chain has a unique steady-state distribution, with $\underline{\mu}(h)>0$ for every $h\in H$. In the high state, $\overline{q}(h)>0$ everywhere, so $\mathbf{1}$ is reachable from every history. Since the Markov chain is finite, this implies that its steady-state distribution is unique \citep[][Theorem 11.2]{StokeyLucas1989}.

Fix an equilibrium $\sigma$. Consider a history ending with $0$, i.e., $h = v0$ for some action sequence $v$ of length $m-1$. The two predecessors of $v0$ are $0v$ and $1v$. Since history probabilities are unchanged in a steady state, $\mu_\theta(v0) = \sum_{x\in\{0,1\}}\mu_\theta(xv)[1-q_\theta(xv)]$. If agents are automatically aware or $v0\neq\mathbf{0}$, then \eqref{eq:order_transition} applies at both predecessor histories. Substituting gives
\begin{align}
z(v0)
&=\delta\sum_{x\in\{0,1\}}[1-\sigma(xv)]z(xv)
 -R_*p\gamma\sum_{x\in\{0,1\}}\underline{\mu}(xv)
<0.\label{eq:order_zero}
\end{align}
Here $1-\overline{q}(xv)=\delta[1-\sigma(xv)]$ and
$1-\underline{q}(xv)=\delta[1-\sigma(xv)]+p\gamma$; the first term is weakly negative by \eqref{eq:order_optimality}, and the second is strictly negative. Thus, whenever agents are automatically aware or $v0 \neq \mathbf{0}$, \eqref{eq:order_optimality} implies $\sigma(v0)=0$. If $v0 = \mathbf{0}$ when agents are initially unaware, $\sigma(\mathbf{0})=0$ by feasibility of $\sigma$. Hence, $\sigma(h)=0$ whenever $h$ ends in $0$.
\end{proof}

\begin{Lemma}\label{lem:order_last_zero}
Fix an equilibrium with $m\ge2$. If $\sigma(h)>0$ for some
$h\neq\mathbf{1}$, replacing the last zero in $h$ by a one gives
a history $h'$ with $l(h')>R_*$ and $\sigma(h')=1$.
\end{Lemma}

\begin{proof}
For an action sequence $w$ of length $k\le m$, a history $h$ ends in $w$ if $h=uw$ for some $u\in\{0,1\}^{m-k}$. We take $\{0,1\}^0=\{\varnothing\}$, so every history ends in the empty sequence. 

Define $P_\theta(w) =\sum_{u\in\{0,1\}^{m-k}}\mu_\theta(uw)$ to be the probability that the observed history ends in $w$. So, $P_\theta(\varnothing)=1$, and $P_\theta(w)=\mu_\theta(w)$ when $k=m$. Also define $T_{\theta}(w)=\sum_{u\in\{0,1\}^{m-k}}\sigma(uw)\mu_\theta(uw)$, which weights the probability of each history $uw$ by the probability of adoption, $\sigma(uw)$.

We have $\underline{T}(w) \geq 0$ and, by \eqref{eq:order_optimality}, $\overline{T}(w)-R_*\underline{T}(w) =\sum_{u\in\{0,1\}^{m-k}}\sigma(uw)z(uw)\ge0$.\footnote{Under initial unawareness, the summand corresponding to $uw = \mathbf{0}$, when present, is zero because
$\sigma(\mathbf{0})=0$.} Moreover, $T_\theta(w0)=0$ whenever $w$ has length less than $m$, by Lemma \ref{lem:order_steady_states}. For any $w$ of length less than $m$, the next history ends in $w1$ precisely when the current history ends in $w$ and the next action is $1$. If agents are automatically aware or $w$ contains a one, the steady-state balance equations and \eqref{eq:order_transition}
therefore give
\begin{align}
P_\theta(w1) &=\alpha_\theta P_\theta(w)+\delta T_\theta(w),
\label{eq:order_block_one}\\
P_\theta(w0) &=(1-\alpha_\theta)P_\theta(w)-\delta T_\theta(w).
\label{eq:order_block_zero}
\end{align}
We treat the remaining sequences under initial unawareness below.

Let $\rho_\theta:=\alpha_\theta/(1-\alpha_\theta)$. For $1 \leq r \leq m-1$ and sequences $v$ of length at most $m-r-1$, define
\begin{align}
D_{\theta,r}(v)
&=P_\theta(v\mathbf{1}_{r+1})
  -\rho_\theta P_\theta(v0\mathbf{1}_r),
\label{eq:order_D}\\
E_{\theta,r}(v)
&=T_\theta(v\mathbf{1}_{r+1})
  -\rho_\theta T_\theta(v0\mathbf{1}_r).
\label{eq:order_E}
\end{align}
Write $\overline{D}_r = D_{\overline{\theta},r}$, with $\underline{D}_r$, $\underline{E}_r$, $\overline{E}_r$ defined similarly. By induction on $r$ we prove $\overline{D}_r(v) \geq R_*\underline{D}_r(v) \geq 0$, for every such $v$. We also show $\overline{E}_r(v) \geq R_*\underline{E}_r(v) \geq 0$ in the same inductive step.

First consider $r=1$ and fix $v$ of length at most $m-2$. When agents are automatically aware, \eqref{eq:order_block_one}--\eqref{eq:order_block_zero} and $T_\theta(v0)=0$ imply
\begin{equation}
D_{\theta,1}(v) =\alpha_\theta[P_\theta(v1)-\rho_\theta P_\theta(v0)]
  +\delta T_\theta(v1) = \delta[\rho_\theta T_\theta(v)+T_\theta(v1)],
\label{eq:order_D_base}
\end{equation}
where the second equality uses $\rho_\theta(1-\alpha_\theta) = \alpha_\theta$. 

When agents are initially unaware, this applies if $v$ contains a one. Otherwise, $v = \mathbf{0}_k$ for some $0 \le k\le m-2$. At $\mathbf{0}$,  we have $q_\theta(\mathbf{0})=\alpha_\theta-(1-p)(1-\gamma)$ and hence
\begin{align*}
P_\theta(v1) &=\alpha_\theta P_\theta(v)+\delta T_\theta(v) -(1-p)(1-\gamma)\mu_\theta(\mathbf{0}),\\
P_\theta(v0) &=(1-\alpha_\theta)P_\theta(v)-\delta T_\theta(v) +(1-p)(1- \gamma)\mu_\theta(\mathbf{0}),\\
P_\theta(v01) &=\alpha_\theta P_\theta(v0) - (1-p)(1-\gamma) \mu_{\theta}(\mathbf{0}),
\end{align*}
where the last identity uses $T_\theta(v0)=0$. Also, $P_{\theta}(v11)=\alpha_\theta P_\theta(v1)+\delta T_\theta(v1)$ because $v1$ contains a one. Substitution into $D_{\theta,1}(v)$ gives \eqref{eq:order_D_base}, as the additional terms cancel because $\rho_\theta(1-\alpha_\theta ) = \alpha_\theta$.

Thus \eqref{eq:order_D_base} holds in both awareness cases. Since $\overline{T}(w)\ge R_*\underline{T}(w)\ge0$ for every $w$ and $\overline{\rho}>\underline{\rho}>0$, it implies $\overline{D}_1(v) \geq R_*\underline{D}_1(v) \geq 0$ for every admissible $v$.

For the inductive step, fix $1 \leq r\le m-1$ and suppose $\overline{D}_r(v) \geq R_*\underline{D}_r(v) \geq 0$ for every $v$ of length at most $m-r-1$. Take $v$ of length $m-r-1$, and let $h=v0\mathbf{1}_r$ and $h'=v\mathbf{1}_{r+1}$. If $\sigma(h)>0$, optimality gives $z(h)\ge0$, and \eqref{eq:order_D} implies
\begin{equation}\label{eq:order_strict_replacement}
z(h') = \overline{\rho}z(h) + R_*(\overline{\rho}-\underline{\rho})\underline{\mu}(h) + \overline{D}_r(v)-R_*\underline{D}_r(v) > 0.
\end{equation}
The inequality is strict because $\overline{\rho} > \underline{\rho}$ and $\underline{\mu}(h) > 0$. Both histories contain a one, so optimality applies in either awareness case. Thus $l(h') > R_*$ and $\sigma(h')=1$.

We next establish $\overline{E}_r(v)\ge R_*\underline{E}_r(v) \geq 0$, which we will use to advance the induction. Let $v$ have length $k \leq m-r-1$. For each $u \in \{0,1\}^{m-k-r-1}$, consider the histories $h = uv0 \mathbf{1}_r$ and $h'=u v \mathbf{1}_{r+1}$. By \eqref{eq:order_E}, $E_{\theta,r}(v)$ is the sum of $\sigma(h')\mu_\theta(h')-\rho_\theta\sigma(h)\mu_\theta(h)$ over all such $u$.

If $\sigma(h)=0$, the summand is $\sigma(h')\mu_\theta(h')$, and optimality gives $\sigma(h')\overline{\mu}(h') \geq R_*\sigma(h')\underline{\mu}(h') \geq 0$. If $\sigma(h)>0$, \eqref{eq:order_strict_replacement} implies $\sigma(h')=1$, so the summand equals $D_{\theta,r}(uv) + \rho_\theta[1-\sigma(h)]\mu_\theta(h)$.
By the induction hypothesis, $\overline{D}_r(uv) \geq R_*\underline{D}_r(uv) \geq 0$.
The second term is nonnegative in the low state and at least $R_*$ times as large in the high state, since $\overline{\mu}(h) \geq R_*\underline{\mu}(h)$ by optimality, $1-\sigma(h) \geq 0$, and $\overline{\rho} > \underline{\rho} > 0$. Summing over all such $u$ therefore gives $\overline{E}_r(v) \geq R_*\underline{E}_r(v) \geq 0$.

Finally, if $r \leq m-2$ and $v$ has length at most $m-r-2$, applying \eqref{eq:order_block_one} to $v\mathbf{1}_{r+1}$ and $v0\mathbf{1}_r$ yields $D_{\theta,r+1}(v) = \alpha_\theta D_{\theta,r}(v)+\delta E_{\theta,r}(v)$. Both sequences contain a one, so this holds in either awareness case. Since $\overline{\alpha} > \underline{\alpha}>0$ and $\delta > 0$,
the inequalities established for $D_r(v)$ and $E_r(v)$ imply $\overline{D}_{r+1}(v)\ge R_*\underline{D}_{r+1}(v) \geq 0$. This proves the inductive step.

By Lemma \ref{lem:order_steady_states}, $\sigma(h)>0$ implies that $h$ ends in $1$. If $h\neq\mathbf{1}$, then $h=v0\mathbf{1}_r$ for some $1 \leq r \leq m-1$ and equation \eqref{eq:order_strict_replacement} proves the lemma.
\end{proof}

To complete the proof of Proposition \ref{Prop:order}, fix an awareness case and an equilibrium. If $m=1$, the result follows from Lemma \ref{lem:order_steady_states}. Assume by way of contradiction $m \geq 2$ and $\sigma(h) > 0$ for some $h \neq \mathbf{1}$. Lemma \ref{lem:order_steady_states} implies that $h$ ends in one. By Lemma \ref{lem:order_last_zero}, replacing its last zero by a one gives a history at which $\sigma = 1$. Repeating this reduces the number of zeros by one and eventually gives $\sigma(\mathbf{1}) = 1$. But then $\overline{q}(\mathbf{1}) = 1$, so the point mass at $\mathbf{1}$ is a high-state steady state. Uniqueness in Lemma \ref{lem:order_steady_states} implies $\overline{\mu}(\mathbf{1})=1$. In particular, $z(h)=-R_*\underline{\mu}(h)<0$, contradicting $\sigma(h) > 0$.

\section{Proof of Proposition \ref{prop:ImperfectInform}}\label{app:ImperfectInform}

We prove the proposition and the equilibrium claims following it. Throughout, $R_*>1$, as in the baseline model. Since $\overline R_a=\overline R_u=1$ when $m=1$, the hypothesis $R_* < \overline{R}_j$ implies $m \geq 2$. Write $\lambda := (1-\varepsilon) / \varepsilon > 1$. Let $\sigma_+(n)$ and $\sigma_-(n)$ denote the probabilities with which informed high-cost agents choose $1$ after signal $+$ and $-$, respectively. Their likelihood ratios are $\lambda l(n)$ and $l(n)/\lambda$, respectively. Thus each chooses $1$ strictly if his likelihood ratio exceeds $R_*$, and $0$ strictly if it is below $R_*$.

Let $q_\theta(n)$ denote the probability of action $1$ conditional on state $\theta$ and history $n$. For $n \geq 1$,
\begin{align*}
\overline{q}(n)
&=1-\gamma + p\gamma[(1-\varepsilon)\sigma_+(n) + \varepsilon \sigma_-(n)] + (1-p)\gamma\sigma(n),\\
\underline{q}(n)
&=1-\gamma+p\gamma[\varepsilon\sigma_+(n) + (1-\varepsilon)\sigma_-(n)] + (1-p)\gamma\sigma(n).
\end{align*}
These expressions apply when $n=0$, $j=a$. When $n=0$, $j=u$, we have $\sigma(0) = 0$ and
\begin{align*}
\overline{q}(0)
&= p(1-\gamma) + p\gamma\big[(1-\varepsilon)\sigma_+(0) + \varepsilon\sigma_-(0)\big], \\
\underline{q}(0)
&= p(1-\gamma) + p\gamma\big[\varepsilon\sigma_+(0) + (1-\varepsilon)\sigma_-(0)\big].
\end{align*}

If $n$ is in the support of both steady-state distributions, we have $\overline{q}(n) \geq \underline{q}(n)$ since $\sigma_+(n) \geq \sigma_-(n)$. Applying the construction in Appendix \ref{secA} to $q_\theta(n)$
gives the unique steady-state distribution
\begin{equation}\label{eq:new-stationary}
\mu_{\theta}(n) = \frac{{m \choose n}
\prod_{i<n}q_\theta(i)\prod_{i>n}[1-q_\theta(i)]}{Z_\theta},
\end{equation}
where $Z_\theta = \sum_{n=0}^m {m \choose n} \prod_{i<n} q_\theta(i) \prod_{i>n} [1-q_\theta(i)]$ is the sum of the numerators. The steady-state balance equations imply that, for $n=0,\ldots,m-1$,
\begin{equation}\label{eq:new-balance}
\mu_\theta(n)q_\theta(n) = \frac{n+1}{m-n}\mu_\theta(n+1)[1-q_\theta(n+1)].
\end{equation}

\begin{Lemma}\label{lem:bothActionspos}
    Both actions occur with positive probability after every history, so $\overline{\mu}$ and $\underline{\mu}$ have full support in equilibrium. 
\end{Lemma}
\begin{proof}
    Since $q_\theta(n)>0$ for all $n$ because of low-cost informed types, it suffices to show that $q_\theta(n)< 1$ for all $n$. Assume by way of contradiction that $k$ is the largest integer at which $q_\theta(k)=1$. By definition of $q_\theta(k)$, we have that $\overline{q}(k) = \underline{q}(k) = 1$ and \eqref{eq:new-balance} implies $\mu_\theta(k-1) = 0$. For every $n<k$, the numerator in \eqref{eq:new-stationary} contains the factor $1-q_\theta(k)=0$, so $\mu_\theta(n) = 0$.
    
    For $n \geq k$, all factors in \eqref{eq:new-stationary} are positive by the definition of $k$, so $\mu_\theta(n) > 0$ for all $n \geq k$. For $n \geq k$, optimality of decisions implies $\sigma_+(n) \geq \sigma_-(n)$ and hence $\overline{q}(n) \geq \underline{q}(n)$. The transitions in each state, given by equation \eqref{eq:new-balance}, imply that, for $m > n \geq k$,
    \[
    \frac{l(n+1)}{l(n)}
    =\frac{\overline q(n)}{\underline q(n)} \frac{1-\underline q(n+1)}{1-\overline q(n+1)} \geq 1.
    \]
    Therefore $l(k) \leq l(n)$ for all $n \geq k$. Since $\sum_{n=k}^{m}\underline\mu(n)l(n) = \sum_{n=k}^{m}\overline\mu(n)=1$, we have $l(k)\le1<R_*$. Uninformed high-cost agents therefore do not adopt at $k$, so $\sigma(k)=0$. But these agents occur with positive probability, which contradicts that $q_\theta(k)=1$.
\end{proof}

\begin{Lemma}\label{lem:rejectBelowHerd}
In every equilibrium, $l(n)<R_*$ and $\sigma(n)=\sigma_-(n)=0$ for all $n<m$.
\end{Lemma}

\begin{proof}
Assume by way of contradiction that $l(n) \geq R_*$ for some $n<m$. At history $n+1 \geq 1$, all agents are aware of action $1$, and low-cost agents always adopt. By Lemma \ref{lem:bothActionspos}, some high-cost agents choose $0$ at history $n+1$, and adding the incentive inequalities for uninformed agents and informed agents receiving either signal gives $\overline{\mu}(n+1)[ 1-\overline{q}(n+1) ] -R_*\underline{\mu}(n+1)[1-\underline{q} (n+1)] \leq 0$. From \eqref{eq:new-balance} we can conclude
\[
\overline{\mu}(n)\overline q(n) - R_*\underline\mu(n)\underline{q}(n) = \frac{n+1}{m-n} \Big[ \overline\mu(n+1)[1-\overline q(n+1)] -R_*\underline\mu(n+1)[1-\underline q(n+1)] \Big] \leq 0.
\]
If $l(n)=R_*$, informed high-cost agents strictly adopt after $s = +$ and strictly reject after $s = -$. Other agents' adoption probabilities are independent of the state, so $\overline q(n)-\underline{q}(n) = p\gamma(1-2\varepsilon) > 0$. Since $\overline{\mu}(n) = R_* \underline{\mu}(n)$, this implies $\overline{\mu}(n) \overline{q}(n) - R_* \underline{\mu}(n) \underline{q}(n) = p \gamma R_* \underline{\mu}(n) (1-2\varepsilon) > 0$, which contradicts the display equation above.

Now consider $l(n) > R_*$. Since informed low-cost agents adopt in both states, their contribution to $\overline\mu(n)\overline{q}(n) - R_* \underline{\mu}(n) \underline{q}(n)$ is therefore $p(1-\gamma) \underline{\mu}(n) [ l(n) - R_*] > 0$.  Thus, $\overline\mu(n)\overline q(n)-R_*\underline\mu(n)\underline q(n) \geq p(1-\gamma)\underline\mu(n)[l(n)-R_*]>0$, contradicting the displayed inequality.

Thus, $l(n)<R_*$ for every $n<m$. Since the likelihood ratio after signal $-$ is $\frac{\varepsilon}{1-\varepsilon} l(n) < R_*$, in an equilibrium we must have $\sigma(n) = \sigma_-(n)=0$ for every $n<m$.
\end{proof}

We now show that informed high-cost agents strictly follow their signals under the assumptions of the Proposition, regardless of how uninformed high-cost agents behave. To that end, fix $x = \sigma(m)\in[0,1]$. By the above lemma, $\sigma(n) = 0$ if $n< m$.

First consider the candidate $\sigma_+(n)=1$, $\sigma_-(n)=0$ at every $n$. 
As in previous proofs, let
\[
\overline{\alpha} = 1-\gamma + p\gamma(1-\varepsilon),
\qquad
\underline{\alpha} = 1-\gamma+p\gamma\varepsilon,
\qquad
\delta=(1-p)\gamma.
\]
Then $q_{\theta}(n)=\alpha_\theta$ for $0<n<m$, and $q_{\theta}(m) = \alpha_\theta + \delta x$. At $n=0$, $q_\theta(0) = \alpha_\theta$ when agents are automatically aware, and $q_\theta(0) = \alpha_\theta-(1-p)(1-\gamma)$ when they are unaware. By \eqref{eq:new-stationary}
\begin{equation}\label{eq:new-endpoints}
\mu_\theta(0;x) = \frac{(1-\alpha_\theta)^{m-1}(1-\alpha_\theta-\delta x)}{Z_\theta(x)}, \qquad
\mu_\theta(m;x) = \frac{q_\theta(0)\alpha_\theta^{m-1}}{Z_\theta(x)}.
\end{equation}
At $x=0$, summing these numerators gives $Z_\theta(0) = (1-\alpha_\theta)^m + \frac{q_\theta(0)}{\alpha_\theta} \big[1-(1-\alpha_\theta)^m\big]$. We also have that $Z_\theta(x) = \frac{1-\alpha_\theta-\delta x}{1-\alpha_\theta}Z_\theta(0) +\frac{\delta x}{1-\alpha_\theta} q_\theta(0) \alpha_\theta^{m-1}$.

If informed high-cost agents follow their signals, $\overline q(n)-\underline q(n)=p\gamma(1-2\varepsilon)>0$ for every $n$. The steady-state balance equations \eqref{eq:new-balance} imply $l(0;x) < \cdots < l(m;x)$. Since likelihood ratios average to one under $\underline{\mu}$, we must also have $l(m;x) > 1$. Differentiating gives
\[
\frac{\partial \log l(0;x)}{\partial x}
=\delta\left[ \frac{\underline\mu(m;x)}{1-\underline\alpha-\delta x} -\frac{ \overline\mu(m;x)} {1-\overline\alpha-\delta x} \right] < 0,
\]
where the inequality holds since $\overline{\mu}(m;x) > \underline{\mu}(m;x)$ and $1 - \overline{\alpha} - \delta x < 1 - \underline{\alpha} - \delta x$. So observing $n=0$ when $x=1$ gives informed high-cost agents the weakest incentive to adopt after signal $+$.

At $x=1$, we have $1-\overline{\alpha} - \delta = p \gamma \varepsilon$, and $1-\underline\alpha-\delta=p\gamma(1-\varepsilon)$. Consequently,
\begin{equation}\label{eq:new-worst-incentive}
\lambda l(0;1)
=\left(\frac{1-\overline\alpha}
{1-\underline\alpha}\right)^{m-1}
\frac{Z_{\underline\theta}(1)}{Z_{\overline\theta}(1)}.
\end{equation}
To evaluate the limit as $\varepsilon \rightarrow 0$, we use the expression for $Z_\theta(x)$ above. When agents are automatically aware,
$q_\theta(0) = \alpha_\theta$ and $Z_\theta(0) = 1$, so $Z_{\overline\theta}(1) \rightarrow [1-(1-p)\gamma]^m$, and $Z_{\underline\theta}(1) \rightarrow p+(1-p)(1-\gamma)^m$. When agents are initially unaware, $q_{\overline\theta}(0) \rightarrow p$, $q_{\underline\theta}(0) \rightarrow p(1-\gamma)$, and $Z_{\underline\theta}(0) \rightarrow p+(1-p)\gamma^m$. Hence $Z_{\overline\theta}(1) \rightarrow p[1-(1-p)\gamma]^{m-1}$ and $Z_{\underline\theta}(1) \rightarrow p \big[p+(1-p)((1-\gamma)^m+\gamma^m)\big]$.  Since $(1-\overline\alpha)/(1-\underline\alpha) \rightarrow 1-p$, substituting into \eqref{eq:new-worst-incentive} gives
$\lim_{\varepsilon\rightarrow 0} \lambda l(0;1)=\overline{R}_j$.

Therefore, if $R_*<\overline R_j$, adoption after signal $+$ is strictly optimal under the candidate equilibrium at every $n$ and every $x\in[0,1]$, for all sufficiently small $\varepsilon$.

By Lemma \ref{lem:rejectBelowHerd}, rejection after signal $-$ is already strictly optimal for informed high-cost agents below the herd. To establish this at the herd, \eqref{eq:new-stationary} gives
\[
\frac1{\mu_\theta(m)} = 1 + \sum_{n<m}{m \choose n} \frac{\prod_{i=n+1}^{m}[1-q_\theta(i)]} {\prod_{i=n}^{m-1}q_\theta(i)}.
\]
Thus the herd probability is weakly increasing in $q_\theta(i)$. Relative to the baseline model at the same $x$, high-state adoption probabilities are weakly lower and low-state adoption probabilities are weakly higher, under any informed strategy. So $\frac{l(m)}{\lambda}
\leq \frac{L_m^j(x)}{\lambda} \leq \frac{\varepsilon}{1-\varepsilon} \max\{L_m^j(0),L_m^j(1)\} \rightarrow 0$, where $L_m^j$ is the baseline herd likelihood ratio. Since the upper bound is uniform over $x$ and informed agent strategies, rejection after $-$ is strictly optimal at the herd in every equilibrium for all sufficiently small $\varepsilon$. Thus informed high-cost agents always follow their signals.

We are left to rule out strategies other than following the signal
for informed high-cost agents in equilibrium. Take $\varepsilon$ small enough that $\lambda>R_*$. In every equilibrium, the preceding bound on $\frac{l(m)}{\lambda}$ gives $\sigma_-(n)=0$ for every $n$. Since $l(m) \geq 1$, we also have $\sigma_+(m) = 1$.

Fix any informed strategy with $\sigma_-(n)=0$ for all $n$ and $\sigma_+(m)=1$. As above, $l(n;x)$ is nondecreasing in $n$, and $l(0;x)$ decreases in $x$: the displayed expression for $\partial \log l(0;x)/\partial x$ remains valid, since under any such strategy $q_\theta(m) = \alpha_\theta + \delta x$ and only $q_\theta(m)$ depends on $x$, and it is negative because $l(m;x) \geq 1$ gives $\overline{\mu}(m;x) \geq \underline{\mu}(m;x)$. Hence $l(n;x) \geq l(0;x) \geq l(0;1)$. At $x=1$, $\lambda[1-\overline q(m)]/[1-\underline q(m)]=1$, so \eqref{eq:new-stationary} implies
\begin{equation}\label{eq:jointcont}
    \lambda l(0;1) = \frac{Z_{\underline\theta}(1)}{Z_{\overline\theta}(1)} \prod_{i=1}^{m-1} \frac{1-\overline q(i)}{1-\underline q(i)}.
\end{equation}

This expression has a finite limit as $\varepsilon \rightarrow 0$. At $\varepsilon=0$, informed high-cost agents in the low state receive $-$ and reject. Thus $Z_{\underline\theta}(1)$ is independent of $\sigma_+$, and $\underline q(i)=1-\gamma$ for $1 \leq i < m$. In the high state, $\overline{q}(m) = 1$, so $Z_{\overline\theta}(1) = \prod_{i=0}^{m-1} \overline{q}(i)$. By substituting we get 
\[
\lim_{\varepsilon \rightarrow 0} \lambda l(0;1) =\left.
\frac{Z_{\underline\theta}(1)}{\gamma^{m-1}} \frac{\prod_{i=1}^{m-1}[1-\overline q(i)]} {\prod_{i=0}^{m-1}\overline q(i)} \right|_{\varepsilon=0}.
\]
At $\varepsilon=0$, increasing any $\sigma_{+}(i)$ for $i<m$ increases $\overline{q}(i)$ without changing $Z_{\underline{\theta}}(1)$. This increases the denominator and weakly decreases the numerator of the limiting expression above. Its minimum is therefore attained where $\sigma_{+}(i)=1$ for every $i$. But this was the equilibrium candidate strategy above, and we showed that it has the limiting likelihood ratio $\overline{R}_{j}$. Consequently, for every informed strategy under consideration, $\lim_{\varepsilon \rightarrow 0} \lambda l(0;1)
\geq \overline{R}_{j}$.

Since the right-hand side of \eqref{eq:jointcont} is continuous in $\varepsilon$ and the informed agents' adoption probabilities (including at $\varepsilon=0$), and its denominators are uniformly positive, convergence is uniform over the compact strategy set $[0,1]^{m}$. Thus $R_{*} < \overline{R}_{j}$ implies, for all sufficiently small $\varepsilon$, $\lambda l(n;x) \geq \lambda l(0;1)>R_{*}$ in every equilibrium. Hence informed high-cost agents strictly adopt after signal $+$ at every history.

\bibliographystyle{ecta}
\bibliography{ref}
\end{document}